%% file: main.tex
\documentclass{aa}  

\usepackage{graphicx}
\usepackage{amsmath,amsfonts,amssymb}
\usepackage{natbib}

\usepackage{txfonts}
\usepackage{xcolor}

\usepackage{blindtext}
\usepackage{float}
\usepackage{dblfloatfix}
\usepackage{afterpage}
\usepackage{ifthen}
\usepackage[morefloats=12]{morefloats}

\usepackage{placeins}
\usepackage{multicol}
\bibpunct{(}{)}{;}{a}{}{,}
\usepackage[switch]{lineno}
\definecolor{linkcolor}{rgb}{0.6,0,0}
\definecolor{citecolor}{rgb}{0,0,0.75}
\definecolor{urlcolor}{rgb}{0.12,0.46,0.7}
\usepackage[breaklinks, colorlinks, urlcolor=urlcolor,
    linkcolor=linkcolor,citecolor=citecolor,pdfencoding=auto]{hyperref}
\hypersetup{linktocpage}
\usepackage{bold-extra}
\usepackage{lipsum}

\input{Planck}

\def\Cosmoglobe{\textsc{Cosmoglobe}}
\def\commander{\texttt{Commander}}

\def\Planck{\textit{Planck}}
\def\Spitzer{\textit{Spitzer}}
\def\Gaia{\textit{Gaia}}

\newcommand{\cii}{\ensuremath{\mathsc {C\ ii}}}

\newcommand{\nWmsr}{\mathrm{nW}\,\mathrm{m}^{-2}\,\mathrm{sr}^{-1}}

\newcommand{\dv}[0]{\vec{d}}
\renewcommand{\t}[0]{\vec{t}}

\newcommand{\B}[0]{\tens{B}}

\newcommand{\G}[0]{\tens{G}}
\newcommand{\n}[0]{\vec{n}}

\newcommand{\s}[0]{\vec{s}}
\renewcommand{\a}[0]{\vec{a}}
\newcommand{\m}[0]{\vec{m}}

\renewcommand{\L}[0]{\tens{L}}

\newcommand{\N}[0]{\tens{N}}
\newcommand{\M}[0]{\tens{M}}

\renewcommand{\r}[0]{\vec{r}}

\renewcommand{\P}[0]{\tens{P}}

\newcommand{\Te}[0]{T_{\rm e}}

\newcommand{\mathsc}[1]{{\normalfont\textsc{#1}}}
\newcommand{\hi}{\ensuremath{\mathsc {H\ i}}}

\newcommand{\cosmoglobe}{\textsc{Cosmoglobe}}

\def\Tcmb{\ifmmode T_\mathrm{CMB}\else $T_{\mathrm{CMB}}$\fi}
\def\Tcold{\ifmmode T_\mathrm{c}\else $T_{\mathrm{c}}$\fi}
\def\Thot{\ifmmode T_\mathrm{h}\else $T_{\mathrm{h}}$\fi}
\def\Tnear{\ifmmode T_\mathrm{n}\else $T_{\mathrm{n}}$\fi}
\def\scmb{\ifmmode s_\mathrm{CMB}\else $s_{\mathrm{CMB}}$\fi}
\def\squad{\ifmmode s_\mathrm{quad}\else $s_{\mathrm{quad}}$\fi}
\def\ssynch{\ifmmode s_\mathrm{s}\else $s_\mathrm{s}$\fi}
\def\sdust{\ifmmode s_\mathrm{d}\else $s_{\mathrm{d}}$\fi}
\def\ssdust{\ifmmode s_\mathrm{sd}\else $s_{\mathrm{sd}}$\fi}
\def\same{\ifmmode s_\mathrm{AME}\else $s_{\mathrm{AME}}$\fi}
\def\ssrc{\ifmmode s_\mathrm{src}\else $s_{\mathrm{src}}$\fi}
\def\sco{\ifmmode s_\mathrm{CO}\else $s_{\mathrm{CO}}$\fi}
\def\sff{\ifmmode s_\mathrm{ff}\else $s_{\mathrm{ff}}$\fi}
\def\gff{\ifmmode g_\mathrm{ff}\else $g_{\mathrm{ff}}$\fi}
\def\fsynch{\ifmmode f_\mathrm{s}\else $f_{\mathrm{s}}$\fi}
\def\fsd{\ifmmode f_\mathrm{sd}\else $f_{\mathrm{sd}}$\fi}
\def\fame{\ifmmode f_\mathrm{AME}\else $f_{\mathrm{AME}}$\fi}
\def\alphasrc{\ifmmode \alpha_\mathrm{src}\else $\alpha_{\mathrm{src}}$\fi}
\def\bcold{\ifmmode \beta_\mathrm{c}\else $\beta_{\mathrm{c}}$\fi}
\def\bhot{\ifmmode \beta_\mathrm{h}\else $\beta_{\mathrm{h}}$\fi}
\def\bnear{\ifmmode \beta_\mathrm{n}\else $\beta_{\mathrm{n}}$\fi}
\def\bsynch{\ifmmode \beta_\mathrm{s}\else $\beta_{\mathrm{s}}$\fi} 
\def\bsun{\ifmmode \beta_\mathrm{sun}\else $\beta_{\mathrm{sun}}$\fi} 
\def\nuzeros{\ifmmode \nu_{0,\mathrm{s}}\else $\nu_{0,\mathrm{s}}$\fi} 
\def\nuzeroff{\ifmmode \nu_{0,\mathrm{ff}}\else $\nu_{0,\mathrm{ff}}$\fi} 
\def\nuzerocold{\ifmmode \nu_{0,\mathrm{c}}\else $\nu_{0,\mathrm{c}}$\fi}
\def\nuzerohot{\ifmmode \nu_{0,\mathrm{h}}\else $\nu_{0,\mathrm{h}}$\fi}
\def\nuzeronear{\ifmmode \nu_{0,\mathrm{n}}\else $\nu_{0,\mathrm{n}}$\fi} 
\def\nuzeroame{\ifmmode \nu_{0,\mathrm{AME}}\else $\nu_{0,\mathrm{AME}}$\fi} 
\def\nuzerosd{\ifmmode \nu_{0,\mathrm{}}\else $\nu_{0,\mathrm{sd}}$\fi} 
\def\nuzerosrc{\ifmmode \nu_{0,\mathrm{src}}\else $\nu_{0,\mathrm{src}}$\fi} 
\def\nup{\ifmmode \nu_{\mathrm{p}}\else $\nu_{\mathrm{p}}$\fi} 
\def\alphasd{\ifmmode \alpha_{\mathrm{sd}}\else $\alpha_{\mathrm{sd}}$\fi} 
\def\Te{\ifmmode T_{\mathrm{e}}\else $T_{\mathrm{e}}$\fi} 
\def\kB{\ifmmode k_\mathrm{B}\else $k_{\mathrm{B}}$\fi}

\begin{document}

\title{\bfseries{\Cosmoglobe\ DR2. II. CIB monopole measurements from \\ COBE-DIRBE through global Bayesian analysis}}

   \input{authors.tex}

   \titlerunning{\Cosmoglobe: CIB monopole constraints}
   \authorrunning{Watts et al.}

   \date{\today}
   
  \abstract{
    We derive new constraints on the Cosmic Infrared Background (CIB) monopole spectrum from a set of reprocessed COBE-DIRBE sky maps that have lower instrumental and astrophysical contamination than the legacy DIRBE maps. These maps have been generated through a global Bayesian analysis framework that simultaneously fits cosmological, astrophysical, and instrumental parameters, as described in a series of papers collectively referred to as \Cosmoglobe\ Data Release~2 (DR2). We have applied this method to the (time-ordered) DIRBE Calibrated Individual Observations (CIO), complemented by selected HFI and FIRAS sky maps to break key astrophysical degeneracies, as well as WISE and \Gaia\ compact object catalogs. In this paper, we focus on the CIB monopole constraints that result from this work. We report positive detections of an isotropic signal in six out of the ten DIRBE bands (1.25, 2.2, 3.5, 100, 140, and 240\,$\mathrm{\mu m}$). For the 2.2\,$\mu$m channel, we find an amplitude of $10.2\pm1.2\,\nWmsr$, which is 74\,\% lower than that reported from the official DIRBE maps. For the 240\,$\mu\mathrm{m}$ channel, we find $6\pm3\,\nWmsr$, which is 56\,\% lower than the official DIRBE release. We interpret these lower values as resulting from improved zodiacal light and Galactic foreground modeling. For the bands between 4.9 and 60\,$\mu\mathrm{m}$, the presence of excess radiation in solar-centric coordinates reported in a companion paper precludes the definition of lower limits. However, the analysis still provides well-defined upper limits. For the 12\,$\mu\mathrm{m}$ channel, we find an upper 95\,\% confidence limit of 55\,$\nWmsr$, more than a factor of eight lower than the corresponding legacy result of 468\,$\nWmsr$. The results presented in this paper redefine the state-of-the-art CIB monopole constraints from COBE-DIRBE, and provide a real-world illustration of the power of global end-to-end analysis of multiple complementary data sets which is the foundational idea of the \cosmoglobe\ project.
  }


   \maketitle

   \setcounter{tocdepth}{2}
   \tableofcontents

\section{Introduction}

The Cosmic Infared Background (CIB) has long been recognized as a key probe of star formation and cosmological large-scale structures \citep{partridge1967}. The unresolved radiation from distant galaxies can be used to probe both the total number density of galaxies at the peak of star formation as well as their clustering properties. Unlike the Cosmic Microwave Background (CMB), whose serendipitous discovery was due to its high photon number density \citep{penzias:1965}, the CIB's total brightness is much fainter, with comparable total brightness to the thermal emission from dust particles in the Milky Way and our own Solar system. 
Searches from balloons (e.g., \citealt{hofmann1978}) and sounding rockets (e.g., \citealt{matsumoto1988} and \citealt{noda1992}) were limited by local sources of radiation, including the atmosphere and rocket exhaust.
As a result, the CIB long eluded direct searches until a series of detector technology breakthroughs in the infrared frequency regime led to its eventual detection. Specifically, the first detection of the CIB monopole was made by \citet{puget1996} by combining COBE-FIRAS data between 400--1000\,$\mathrm{\mu m}$ with \hi\ measurements to subtract Galactic dust emission. These measurements were later independently confirmed through various other techniques and data sets \citep[e.g.,][]{fixsen1998, schlegel1998, lagache:1999, matsumoto:2005, matsuura:2011, penin:2012, tsumura:2013, matsuura:2017}.

The first satellite instrument that was specifically designed to detect and characterize both the CIB monopole and fluctuations was the Diffuse Infrared Background Experiment (DIRBE; \citealp{hauser1998}), which flew on-board the NASA-led Cosmic Background Explorer (COBE) satellite. DIRBE continuously observed the infrared sky in ten wavelength bands between 1.25 and 240\,$\mu\mathrm{m}$ for about 10 months. In retrospect, and according to subsequent measurements by later experiments such as \Spitzer\ \citep{dole:2006} and \Planck\ \citep[e.g.,][]{planck2013-pip56,planck2016-XLVIII}, DIRBE did in fact have the raw sensitivity that was required to make a definitive measurement of both the CIB monopole and anisotropies \citep{boggess92,hauser1998}. However, local astrophysical emission in the form of starlight and thermal dust emission from the Milky Way and zodiacal dust emission from the Solar system proved to be an enormous modeling challenge. At the same time, these data opened up an entirely new window for modeling these phenomena, and great efforts were spent on establishing increasingly accurate models. For instance, by establishing a three-dimensional model of dust particles in the Solar system, \citet{kelsall1998} derived a ground-breaking model of zodiacal light (referred to as K98) from the time-ordered DIRBE data that to this date serve as a standard reference. Similarly, using the DIRBE $100\,\mathrm{\mu m}$ frequency map as a template, \citet{arendt1998} modeled thermal dust emission in the Milky Way to high enough precision that \citet{hauser1998} could robustly report measurements of the CIB monopole at 140 and $240\,\mathrm{\mu m}$. For short wavelengths between 1 and 25$\,\mu\mathrm{m}$, \citet{arendt1998} established a model for starlight emission that led to unprecedented upper limits on the CIB monopole spectrum. These analyses were quickly followed up by many other authors who leveraged supplementary data sets, turning the original upper limits into positive detections at both 2.2 $\mathrm{\mu m}$ \citep{wright:2000,gorjian:2000,wright:2001} and $3.5\,\mathrm{\mu m}$ \citep{dwek:1998b,gorjian:2000,wright:2000}. Other contemporary analyses were performed, but the results were never fully corroborated and confirmed as fully isotropic signals; see, e.g., \citet{hauser:2001} for a comprehensive review of these early efforts. A recent analysis by \citet{sano:2020} utilized the DIRBE Solar elongation maps to disentangle the isotropic zodiacal emission from the CIB, highlighting the DIRBE dataset's continued relevance in the search for the CIB.

Despite these massive efforts in terms of astrophysical modeling, the final foreground residuals turned out to be too large to allow robust CIB measurements across the full DIRBE wavelength range. For instance, while the K98 model was a huge leap forward in terms of understanding the zodiacal light emission, it still only had a $\sim$\,$99\,\%$ accuracy. Given that the absolute level of zodiacal light emission at 25$\,\mu\mathrm{m}$ is 60\,$\mathrm{MJy\,sr^{-1}}$, the corresponding monopole upper limit reported by \citet{hauser1998} was 0.5\,$\mathrm{MJy\,sr^{-1}}$, two orders of magnitude higher than the theoretically predicted CIB monopole. Similarly, residual starlight emission precluded a definitive detection in the near-infrared regime, while residual thermal dust emission from the Milky Way resulted in large uncertainties in the far infrared regime. 

Since the DIRBE data were made public in 1996, many other experiments have dramatically improved our knowledge of the $4\pi$ infrared sky, including WISE \citep{wright:2010}, \Planck\ HFI \citep{planck2016-l03}, and \Gaia\ \citep{gaia:2016}. Each of these provide key ancillary information that may be useful to extract CIB information from  DIRBE. For instance, \Planck\ provides an unprecedented view of Galactic thermal dust, both in terms of sensitivity and angular resolution; WISE provides the location and normalization of Galactic compact objects in the same infrared wavelengths as DIRBE; while \Gaia\ provides a detailed parameter library which can be used to model the spectral energy distribution (SED) of many of the stars in their catalogue, and either through direct use of these models, or reasonable extrapolation to similar stars, the SED of virtually every star measured by DIRBE can be modeled \citep{CG02_04}.

Not only has there been made great observational breakthroughs since the time of the DIRBE release, but this progress has also been accompanied closely by corresponding efforts in algorithm development. One striking example of this is the field of cosmic microwave background (CMB) analysis, which during more than three decades have established a wide range of highly efficient tools to search for weak signals in data that are contaminated by instrumental noise, astrophysical foregrounds, and systematics \citep[e.g.,][]{bennett2012,pb2015,planck2016-l01}. One specific branch of this community-wide effort focused on so-called global Bayesian analysis \citep{jewell2004,wandelt2004}, ultimately culminating in the \cosmoglobe\footnote{\url{https://cosmoglobe.uio.no}}  project which aims to create a single coherent sky model from as many complementary state-of-the-art experiments as possible. The first application of this framework was described in \cosmoglobe\ Data Release 1 (DR1; \citealp{watts2023_dr1}), which performed the first joint analysis of time-ordered data from both the \textit{Wilkinson} Microwave Anisotropy Probe (WMAP; \citealp{bennett2012}) and \Planck\ LFI \citep{planck2016-l02,bp01}. 

In the current \cosmoglobe\ DR2 data release, we apply the same process to the DIRBE data with the goal of characterizing the CIB monopole spectrum from 1 to 240$\,\mu\mathrm{m}$. Even after this work, several key problems remain that will require further investigation, both in terms of astrophysical and instrumental modeling. The following results therefore are not the definitive results that ultimately can be derived from DIRBE. Rather, they are the first step in a long process, in which the CIB is gradually being harnessed through the same Bayesian end-to-end methods that have proven effective for analyzing CMB experiments, with the ultimate goal of seamlessly merging the two fields. Two natural next steps in this process are, first, to include high-resolution experiments such as IRAS \citep{neugebauer:1984} and AKARI \citep{murakami:2007} that can break important degeneracies in the current zodiacal light model, and, second, to reprocess the \Planck\ HFI measurements at the level of raw TOD; preliminary steps towards both of these projects have recently been made, and volunteers are welcome to join this Open Source effort.

The rest of the paper is organized as follows.
We start by briefly reviewing the overall \cosmoglobe\ DR2 data processing algorithm in Sect.~\ref{sec:algorithm} and the data in Sect.~\ref{sec:data}. The corresponding CIB monopole constraints are discussed in Sect.~\ref{sec:results}. We conclude in Sect.~\ref{sec:conclusions}, where we also discuss both potential future improvements and the impact of external data.

\section{Algorithms}
\label{sec:algorithm}

The main data set used in this paper is the zodiacal light subtracted
DIRBE frequency maps described by \citet{CG02_01}. In this section, we
briefly review the algorithm that was used to generate these maps.

The first task in any end-to-end Bayesian \cosmoglobe-style analysis
is to write down an explicit parametric model for the raw time-ordered
data. For the current DIRBE-targeted analysis, we adopt the following
model \citep{CG02_01},
\begin{align}
	\label{eq:model}
	\dv &=\G\P\left[\B\sum_{c=1}^{n_{\mathrm{comp}}}\M_c\a_c+\s_{\mathrm{zodi}} +
          \s_{\mathrm{static}}\right] + \n,\\
        &\equiv \s^{\mathrm{tot}} + \n.
\end{align}
In this expression, $\dv$ denotes a stacked vector of the DIRBE
Calibrated Individual Observations (CIO) for each frequency band; in
the following, we will refer to these as time-ordered data (TOD),
following the CMB nomenclature. Next, $\G$ is in principle an
$n_{\mathrm{tod}}\times n_{\mathrm{tod}}$ diagonal matrix with an
overall constant gain calibration factor per frequency channel; unless
otherwise noted, this is usually set to unity, and we mostly adopt the
official DIRBE calibration as is.\footnote{In principle, one could
marginalize over the gain uncertainties quoted by the DIRBE team
within the Gibbs sampler, but this would increase the Markov chain
correlation length significantly. In practice, it is computationally
more convenient to propagate these uncertainties analytically into the
final results.}  Moving on, $\P$ denotes the satellite pointing
matrix defined in Galactic coordinates, and $\B$ is an
instrumental beam convolution operator. Neither of these are
associated with any free parameters, but are adopted as perfectly
known quantities. The sum runs over $n_{\mathrm{comp}}$ astrophysical
components, each described by a free amplitude $\a_c$ and a mixing
matrix, $\M_c$, the latter of which is defined by some small number of
spectral parameters, $\beta_{\mathrm{c}}$; for full details, see
the companion papers \citet{CG02_04} and \citet{CG02_05}. Next, $\s_{\mathrm{zodi}}$ denotes zodiacal
light emission, which in the current model is described by $\sim$\,70
free parameters collectively denoted $\zeta_{\mathrm{z}}$
\citep{CG02_02}. The static component described by
$\s_{\mathrm{static}}$ is extensively discussed by \citet{CG02_01},
and we will return to this term in Sect.~\ref{sec:excessrad}; for now,
we note that it is described by an amplitude per pixel in
solar-centric coordinates, $\a_{\mathrm{static}}$. Finally, $\n$
is Gaussian instrumental noise with free parameters
$\xi_{\mathrm{n}}$, the most important of which are the white noise
rms's per time sample for each band.

To complete the data model, we need to specify the astrophysical sky
contribution. In the current analysis, we adopt the following sky model
\begin{alignat}{4}
  \sum_{c=1}^{n_{\mathrm{comp}}} \M_c \a_c  = \,
  &\M_{\mathrm{mbb}}(\bcold,\Tcold,\nuzerocold)\vec{a}_{\mathrm{cold}}
  && \textrm{(Cold dust)}\label{eq:skymodel}\\
  + &\M_{\mathrm{mbb}}(\bhot,\Thot,\nuzerohot)
  \vec{a}_{\mathrm{hot}} && \textrm{(Hot dust)}\nonumber \\
  + &\M_{\mathrm{mbb}}(\bnear,\Tnear,\nuzeronear) \t_{\mathrm{near}}
  a_{\nu} && \textrm{(Nearby dust)} \nonumber \\
  + &\left(\frac{\nuzeroff}{\nu}\right)^2
  \frac{g_{\mathrm{ff}}(\nu;\Te) }{g_{\mathrm{ff}}(\nuzeroff;\Te)}
  \vec{t}_{\mathrm{ff}} && \textrm{(Free-free)} \nonumber\\
  + &\delta(\nu-\nu_{0,\mathrm{CO}}^i) \t_{\mathrm{CO}}
  h^{\mathrm{CO}}_{\nu,i} && \textrm{(CO)}\nonumber\\
  + &\delta(\nu-\nu_{0,\cii}) \a_{\cii}
	h^{\cii}_{\nu} && (\cii)\nonumber \\
  + &U_{\mathrm{mJy}} \sum_{j=1}^{n_{\mathrm{s}}}
  f_{\mathit{Gaia},j} a_{\mathrm{s},j}, &\quad&
  \textrm{(Bright stars)} \nonumber\\
  + &U_{\mathrm{mJy}} f_{\mathit{Gaia},j} \a_{\mathrm{fs},j}, &\quad&
  \textrm{(Faint stars)} \nonumber\\  
    + &U_{\mathrm{mJy}} \sum_{j=1}^{n_{\mathrm{e}}}
  M_{\mathrm{mbb}}(\beta_{\mathrm{e},j},
  T_{\mathrm{e},j}, \nu_{0,\mathrm{e}})
  a_{\mathrm{e},j} && \textrm{(FIR sources)}\nonumber\\
  + &m_{\nu} && \textrm{(Monopole)}, \nonumber
\end{alignat}
where rows describe respectively, from top to bottom, 1) cold dust
emission; 2) hot dust emission; 3) nearby dust emission; 4) free-free
emission; 5) CO emission; 6) \cii\ emission; 7) bright starlight
emission; 8) faint starlight emission; 9) other compact sources; and
10) one monopole for each band. Collectively, we define
$\a_{\mathrm{sky}}$ to be the set of all signal amplitude maps, and
$\beta_{\mathrm{sky}}$ to be the set of all SED parameters. The
specific meaning of each symbol in Eq.~\eqref{eq:skymodel} is not
relevant for the present paper, so we
refer the interested reader to \citet{CG02_04} and \citet{CG02_05} for
full details.

Finally, we define $\omega = \{\G,\xi_{\mathrm{n}},
\beta_{\mathrm{sky}},\a_{\mathrm{sky}},\zeta_{\mathrm{z}},\a_{\mathrm{static}}\}$
to be the set of all free parameters in the model. Since this model involves
a free amplitude per pixel for each foreground model, over
hundreds of millions of parameters are being fitted simultaneously,
all of which interact non-trivially.

The \cosmoglobe\ framework is designed to perform classical Bayesian
parameter estimation with a data model such as Eq.~\eqref{eq:model},
and the main goal is then the full joint posterior distribution,
$P(\omega\mid\dv)$, which is given by Bayes' theorem,
\begin{equation}
P(\omega\mid\dv) = \frac{P(\dv\mid\omega) P(\omega)}{P(\dv)} \propto
\mathcal{L}(\omega) P(\omega).
\end{equation}
Here $P(\dv\mid\omega) = \mathcal{L}(\omega)$ is called the likelihood,
$P(\omega)$ denotes a set of priors, and $P(\dv)$ is called the evidence,
which for our parameter estimation purposes is a 
normalization constant. The likelihood is defined by the key assumption that
the instrumental noise is Gaussian distributed,
\begin{equation}
-2\ln\mathcal{L}(\omega) = (\dv-\s^{\mathrm{tot}}(\omega))^t
  \N_{\mathrm{w}}^{-1}(\dv-\s^{\mathrm{tot}}(\omega)) \equiv \chi^2(\omega),
\end{equation}
once again up to a normalization constant. The priors
adopted in the \cosmoglobe\ DR2 analysis are described by
\citet{CG02_01}, \citet{CG02_02}, and \citet{CG02_04}, but for this particular paper the
most important ones are simply that we assume the CIB and the
static components both to be strictly positive. 

With these definitions ready at hand, we employ a standard Monte Carlo
algorithm called Gibbs sampling \citep[e.g.,][]{geman:1984} to map out
the joint posterior distribution, as implemented in a computer code
called \commander\ \citep{eriksen:2004,seljebotn:2019,bp03}. This has
already been applied successfully to a wide range of CMB data sets
including \Planck\ \citep{planck2014-a12,bp01} and WMAP
\citep{watts2023_dr1}. The current analysis is, however, its first
application to infrared wavelengths.

Rather than drawing samples
directly from a large and complicated joint distribution, in Gibbs sampling one
draws samples iteratively from each conditional distribution. For the
data model described above, this translates into the following
so-called Gibbs chain:
\begin{alignat}{11}
\G &\,\leftarrow P(\G&\,\mid &\,\dv,&\, &\,\phantom{\G} &\,\xi_n, &
\,\beta_{\mathrm{sky}}& \,\a_{\mathrm{sky}}, &\,\zeta_{\mathrm{z}},
&\,\a_{\mathrm{static}})\label{eq:gibbs_G}\\
\xi_{\mathrm{n}} &\,\leftarrow P(\xi_{\mathrm{n}}&\,\mid &\,\dv,&\, &\,\G, &\,\phantom{\xi_n} &
\,\beta_{\mathrm{sky}}& \,\a_{\mathrm{sky}}, &\,\zeta_{\mathrm{z}},
&\,\a_{\mathrm{static}})\\
\beta_{\mathrm{sky}} &\,\leftarrow P(\beta_{\mathrm{sky}}&\,\mid &\,\dv,&\, &\,\G, &\,\xi_n, &
\,\phantom{\beta_{\mathrm{sky}}}& \,\a_{\mathrm{sky}}, &\,\zeta_{\mathrm{z}}, &\,\a_{\mathrm{static}})\\
\a_{\mathrm{sky}} &\,\leftarrow P(\a_{\mathrm{sky}}&\,\mid &\,\dv,&\, &\,\G, &\,\xi_n, &
\,\beta_{\mathrm{sky}},& \,\phantom{\a_{\mathrm{sky}},}
&\,\zeta_{\mathrm{z}}, &\,\a_{\mathrm{static}})\\
\zeta_{\mathrm{z}} &\,\leftarrow P(\zeta_{\mathrm{z}}&\,\mid &\,\dv,&\, &\,\G, &\,\xi_n, &
\,\beta_{\mathrm{sky}},& \,\a_{\mathrm{sky}},
&\,\phantom{\zeta_{\mathrm{z}},} &\,\a_{\mathrm{static}})\label{eq:gibbs_zodi}\\
\a_{\mathrm{static}} &\,\leftarrow P(\a_{\mathrm{static}}&\,\mid &\,\dv,&\, &\,\G, &\,\xi_n, &
\,\beta_{\mathrm{sky}},& \,\a_{\mathrm{sky}}, &\,\zeta_{\mathrm{z}} &\,\phantom{\a_{\mathrm{static}}})\label{eq:gibbs_static}.
\end{alignat}
Here, the symbol $\leftarrow$ indicates the operation of drawing a
sample from the distribution on the right-hand side. After some
burn-in period, the resulting joint parameter sets will correspond to
samples drawn from the true underlying joint posterior.

Each sampling step in this algorithm is described by \citet{CG02_01}
and references therein. In some cases, important 
approximations are made that strictly speaking violate the Gibbs
rule, either with the goal of increasing robustness with respect to
systematic errors at the cost of increased statistical uncertainties,
or simply for algorithmic robustness. A prime example of the former is
the fact that our simple model discussed above is not able to fully
describe the Galactic plane, and we therefore apply different
confidence masks for different applications. An important example of
the latter is the zodiacal light model, whose posterior distribution
exhibits a large number of local minima due to strong internal
parameter degeneracies, and a strict Gibbs sampling algorithm may
easily become trapped. For this particular sampling step, we have
therefore instead opted for a simple non-linear Powell algorithm that
is initialized some random parameter distance away from the previous
sample, and then searches for the local minimum. This algorithm is
able to escape local minima, but it comes at the cost of larger
uncertainties than what would result from an ideal posterior mapper.

\section{CIB residual maps and confidence masks}
\label{sec:data}

In this section we give a brief overview of the input datasets that
are used as inputs in the \cosmoglobe\ DR2 analysis, as well as the
output residual maps that serve as the main CIB monopole tracer in the
current paper. For full details, we refer the interested reader to
\citet{CG02_01}. 

\subsection{Data overview}
\label{sec:datasummary}

The calibrated DIRBE TOD forms the primary dataset of interest for
\cosmoglobe\ DR2. In total, there are 285 days of time-ordered
observations at each frequency band, and the total compressed DIRBE
data volume is 20\,GB. The angular resolution of each band is
$42\arcm$ FWHM due to a common optical design with a field stop for straylight removal, resulting in  only small variations between detectors.

The model described by
Eqs.~\eqref{eq:model}--\eqref{eq:skymodel} is very rich, and exhibits
many strong degeneracies of both instrumental and astrophysical
origin. If we were to fit this model to the DIRBE TOD alone, the final
solution would become strongly degenerate. To break these
degeneracies, we include four other complementary main data sets as
part of $\dv$ \citep{CG02_01}.

\paragraph{\Planck\ HFI:} We include one \Planck\ HFI PR4 \citep{npipe} detector sky map for each of the 100, 143, 217,
  353, 545, and 857\,GHz frequency bands to constrain the morphology
  of thermal dust emission. To ensure that neither CMB nor CIB
  fluctuations from these channels contaminate potential CIB results
  from DIRBE, we subtract the \commander\ PR4 CMB temperature map
  from all channels, as well as the \Planck\ PR3 GNILC
  CIB fluctuation maps \citep{gnilc_cib} for the 353, 545, and 857\,GHz channels; any
  residual CMB or CIB fluctuations that may remain after these
  corrections due to modeling uncertainties are much smaller than
  both thermal dust emission and instrumental noise. The angular
  resolutions of these sky maps vary between 5 and $10\arcm$.

  \paragraph{WISE+\textit{Gaia}:} The dominant emission mechanism on short
  wavelengths is starlight radiation, and many methods have been used
  to model this in the literature to date. The original DIRBE analysis
  by \citet{arendt1998} adopted a phenomenological DIRBE-oriented
  model that focused on the overall large-scale morphology, while for
  instance \citet{wright:2001} used 2MASS as a starlight tracer. Since
  that time, WISE \citep{wright:2010} and \Gaia\ \citep{gaia:2016} have revolutionized our understanding of
  starlight emission, and these datasets form the basis of the
  \cosmoglobe\ DR2 model described by \citet{CG02_04}.

  \paragraph{COBE-FIRAS:} Finally, we also include a subset of the
  COBE-FIRAS \citep{mather:1994} sky maps in the current analysis, for two main
  reasons. First, FIRAS serves as a powerful validation source for
  the absolute calibration of the DIRBE 140 and 240\,$\mu\mathrm{m}$
  frequency channel. Second, there is a
  strong emission line present in the DIRBE 140\,$\mu\mathrm{m}$
  channel at 158$\,\mu\mathrm{m}$ due to \cii. By combining high
  spectral resolution information from FIRAS with high spatial
  resolution information from DIRBE 140\,$\mu\mathrm{m}$, a novel
  full-sky \cii\ template is derived as part of the current data
  release. This component has to our knowledge never been accounted
  for in previous CIB studies of the DIRBE 140\,$\mu\mathrm{m}$ channel.

The computational cost for producing one single Gibbs sample according
to the algorithm described in Sect.~\ref{sec:algorithm} is about
500\,CPU-hrs or 4~wall-hours when run on a 128-core cluster node. In
total, we have produced 930 samples for the current analysis,
parallelized over six chains, for a total computational cost of
470k\,CPU-hrs. The first 20~samples are removed as burn-in from each
chain \citep{CG02_02}, leaving a total of 810~samples for scientific
analysis. The total wall time was one month.

\subsection{Implications of excess static radiation}
\label{sec:excessrad}

The parametric model described in Sect.~\ref{sec:algorithm} includes a
component denoted $\s_{\mathrm{static}}$ which describes a
contribution from excess radiation that appears stationary in
solar-centric coordinates. The existence of such excess radiation in
the DIRBE data was reported already by \citet{leinert:1998}, and
illustrated in their Figure~54. The first detailed and systematic
study of this effect, however, has only now been performed as part of
\cosmoglobe\ DR2 \citep{CG02_01}. Since this component plays a
particularly critical role in constraining the CIB monopole with DIRBE
data, we provide a brief review of the effect here.

\begin{figure}
  \centering
  \includegraphics[width=\linewidth]{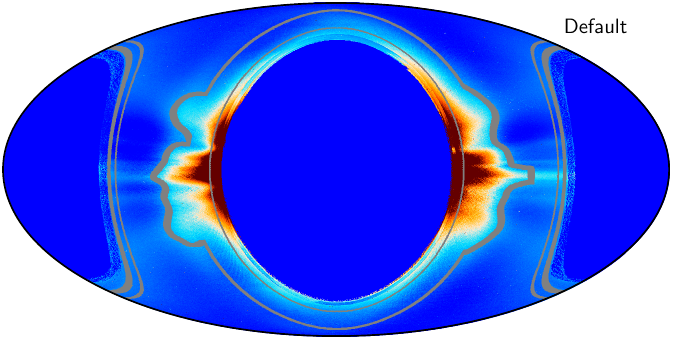}\\
  \includegraphics[width=\linewidth]{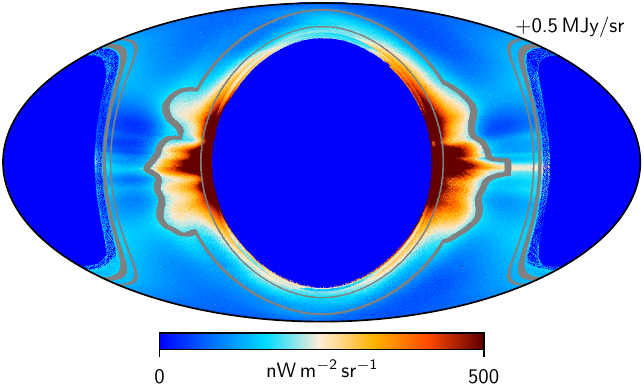}
	\caption{Solar-centered maps derived from residual CIO's at $25\,\mathrm{\mu m}$. (\emph{Top:}) Default excess radiation model, $\a_{\mathrm{static}}$, for the 25\,$\mu$m channel, plotted in solar-centric coordinates; see \citet{CG02_01} for full discussion. Dark blue pixels are never observed by the DIRBE instrument. Thick gray lines indicate the excess radiation mask used in the \cosmoglobe\ DR2 analysis, and thin gray lines show the solar elongation mask used in the legacy DIRBE analysis of $e < 68^{\circ}$ and $e >120^{\circ}$ \citep{kelsall1998}. (\emph{Bottom:}) Same as above, but with an additional offset of +0.5\,$\mathrm{MJy\,sr^{-1}}$, corresponding to the measured monopole at 25\,$\mu$m reported in this paper.}
  \label{fig:sidelobe}
\end{figure}

Excess radiation that appears stationary in solar-centric coordinates
could arise from at least two physical sources. First, any zodiacal
light component that actually is stationary in solar-centric
coordinates could obviously be described by this term. Two well-known
examples are the so-called ``circumsolar ring'' and ``Earth-trailing
feature'' in the K98 model \citep{kelsall1998}, which originate from
dust particles trapped in the joint gravitational field of the
Earth-Sun system. Another possible source is straylight contamination
from the Sun due to telescope non-idealities. As discussed by
\citet{hauser1998}, the DIRBE optics were specifically designed to
minimize such contamination, and no corrections were made for
straylight radiation in the legacy analysis. 

Whatever the origin of the excess radiation might be, \citet{CG02_01}
have now mapped its spatial structure and amplitude for each DIRBE
frequency band by binning the residual TOD, $\dv - \s^{\mathrm{tot}}$,
into solar-centric coordinates. The strongest signal is found at the
25\,$\mu\mathrm{m}$ channel, and this is reproduced in the top panel
of Fig.~\ref{fig:sidelobe}. In this figure, the Sun is located in the
center, and the equator is aligned with the Ecliptic plane. Dark blue
pixels indicate directions that are not observed by the DIRBE
instrument. For reference, \citet{kelsall1998} noted that their
zodiacal light model showed significant residuals for solar elongation
angles smaller than $68^{\circ}$ and larger than $125^{\circ}$, and
those observations were excluded from their zodiacal light mission
average (ZSMA) maps; those limits are marked as gray lines in
Fig.~\ref{fig:sidelobe}. Based on the updated analysis of
\citet{CG02_01}, it is now in retrospect clear that all the excess
signal seen between the two gray lines is in fact present in the final
legacy DIRBE ZSMA maps, and has affected all scientific results
derived from those sky maps during the last three decades.

Qualitatively similar signals were found at all bands between 4.9 and
$100\,\mu\mathrm{m}$, while for wavelengths between 1.25 and
3.5$\,\mu\mathrm{m}$ only weak signatures are visible. At the two
longer wavelengths, 140 and $240\,\mu\mathrm{m}$, there is no evidence
for excess radiation at all. In this respect it is worth noting that
the ten DIRBE detectors are divided into three groups in the optical
path of the instrument \citep{silverberg93}, with the same
grouping as observed for the strength of the solar-centric excess
signal.

As far as the current paper is concerned, it is irrelevant whether
this excess signal is due to a yet-unknown zodiacal light component or
straylight from the Sun. Given that its amplitude reaches 5\,$\mathrm{MJy\,sr^{-1}}$,
it must in either case be fitted and removed from the raw time-ordered
data prior to mapmaking. Given the complex structures seen in
Fig.~\ref{fig:sidelobe}, it does not appear satisfactory simply to exclude
a fixed range of solar elongations from the final mapmaking, given
that the excess reaches several $\mathrm{MJy\,sr^{-1}}$ even at moderate solar
elongations.

  \begin{figure*}
  \centering
  \includegraphics[width=0.42\linewidth]{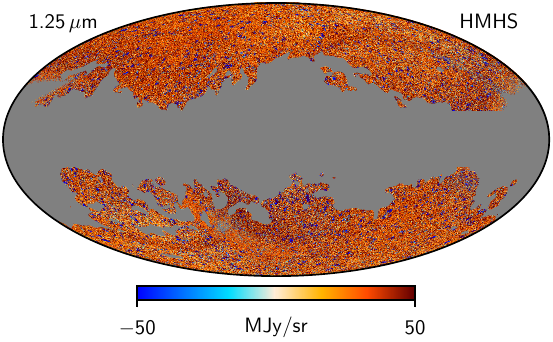}\hspace*{5mm}
  \includegraphics[width=0.42\linewidth]{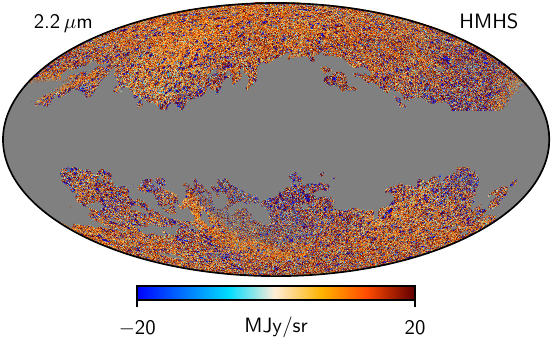}\\
  \includegraphics[width=0.42\linewidth]{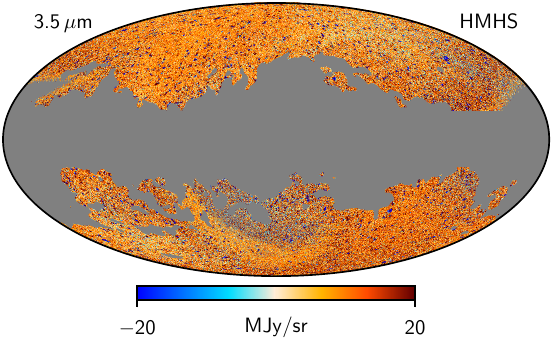}\hspace*{5mm}
  \includegraphics[width=0.42\linewidth]{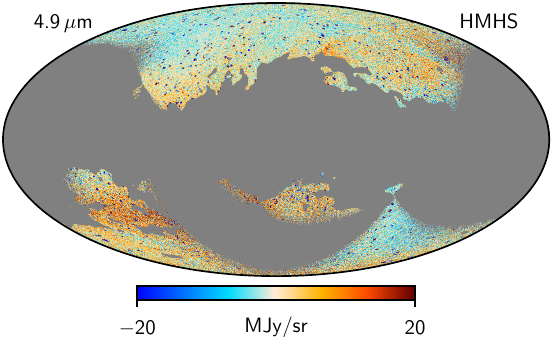}\\
  \includegraphics[width=0.42\linewidth]{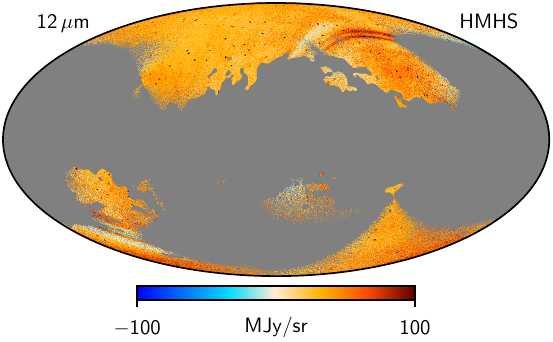}\hspace*{5mm}
  \includegraphics[width=0.42\linewidth]{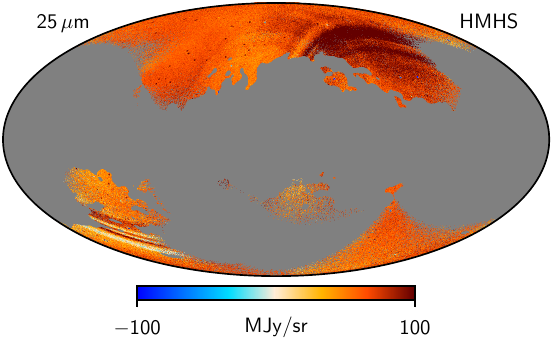}\\
  \includegraphics[width=0.42\linewidth]{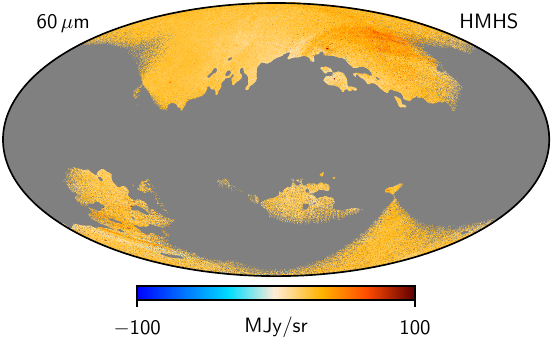}\hspace*{5mm}
  \includegraphics[width=0.42\linewidth]{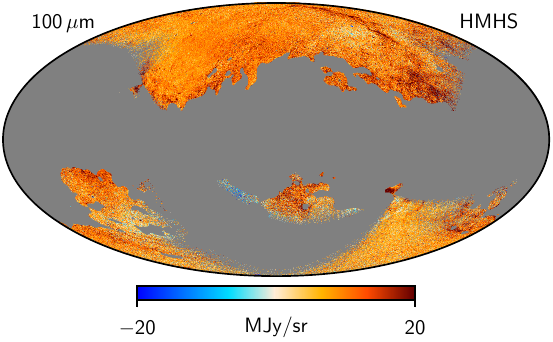}\\
  \includegraphics[width=0.42\linewidth]{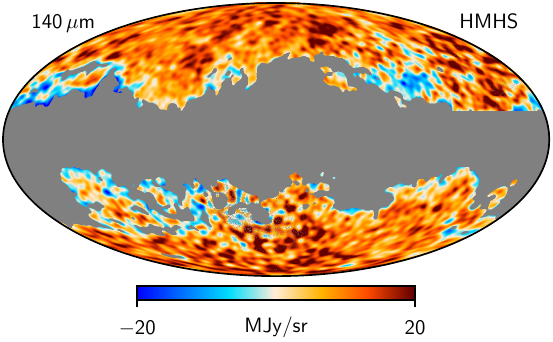}\hspace*{5mm}
  \includegraphics[width=0.42\linewidth]{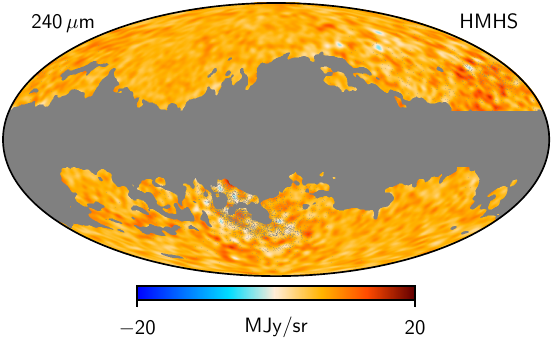}
	  \caption{Half-mission half-sum maps, $(\m_{\mathrm{HM1}}+\m_{\mathrm{HM2}})/2$ for each DIRBE frequency channel. Gray pixels indicate the union of a Galactic mask and the requirement that any pixels must be observed during both HM1 and HM2. The 140 and 240\,$\mu\mathrm{m}$ maps have been smoothed to an angular resolution of $3^{\circ}$ FWHM.}
  \label{fig:hmhs}
\end{figure*}

\begin{figure*}
  \centering
  \includegraphics[width=0.42\linewidth]{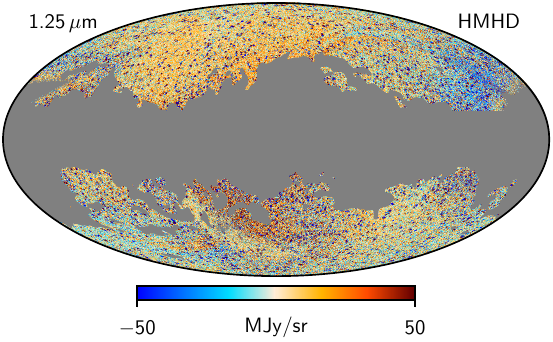}\hspace*{5mm}
  \includegraphics[width=0.42\linewidth]{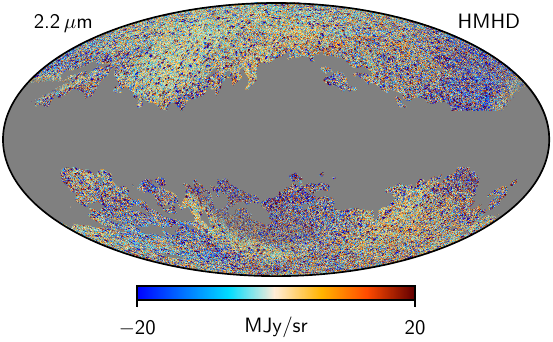}\\
  \includegraphics[width=0.42\linewidth]{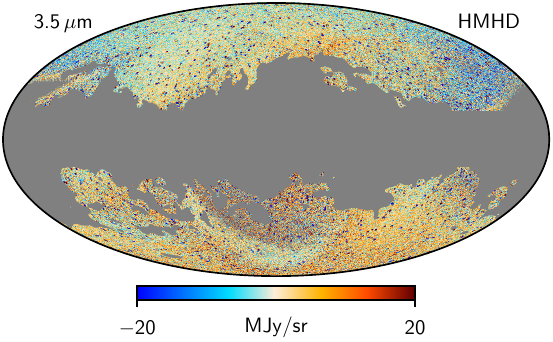}\hspace*{5mm}
  \includegraphics[width=0.42\linewidth]{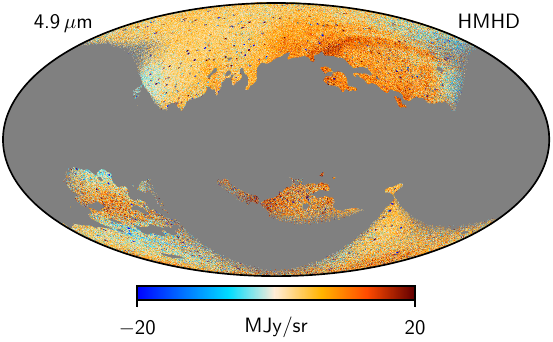}\\
  \includegraphics[width=0.42\linewidth]{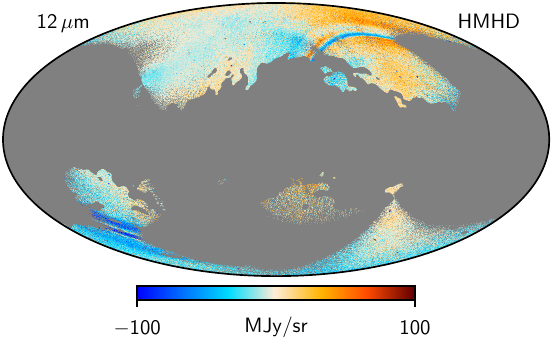}\hspace*{5mm}
  \includegraphics[width=0.42\linewidth]{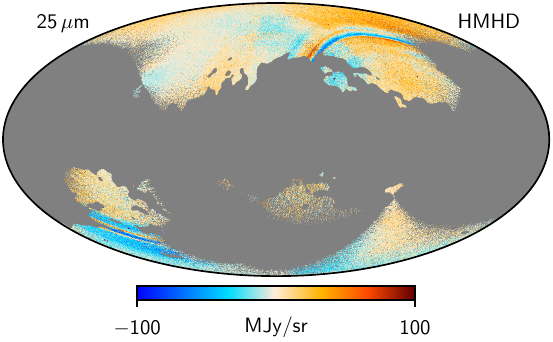}\\
  \includegraphics[width=0.42\linewidth]{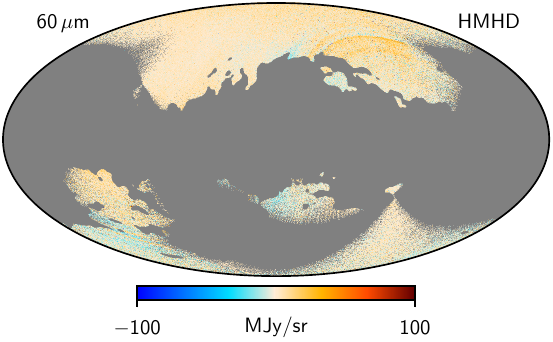}\hspace*{5mm}
  \includegraphics[width=0.42\linewidth]{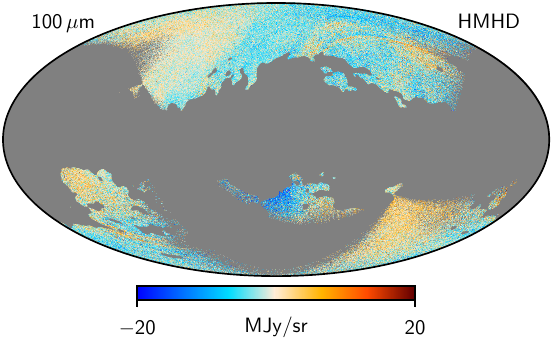}\\
  \includegraphics[width=0.42\linewidth]{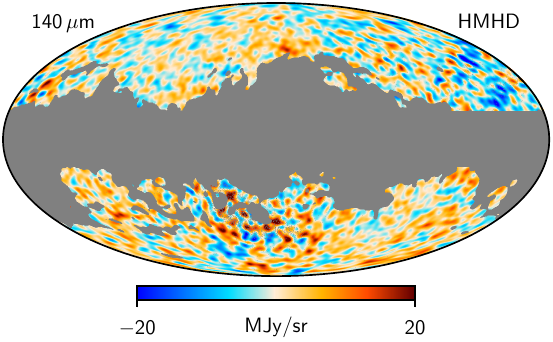}\hspace*{5mm}
  \includegraphics[width=0.42\linewidth]{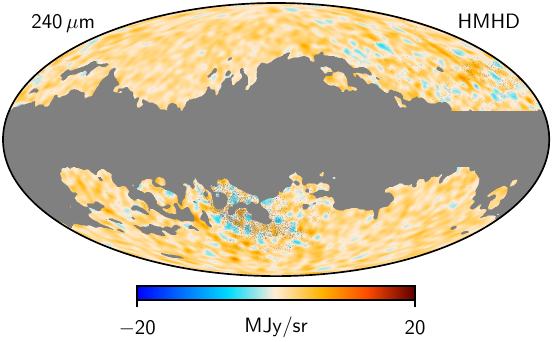}
  \caption{Half-mission half-difference maps, $(\m_{\mathrm{HM1}}-\m_{\mathrm{HM2}})/2$ for each DIRBE frequency channel. Gray pixels indicate the union of a Galactic mask and the requirement that any pixels must be observed during both HM1 and HM2. The 140 and 240\,$\mu\mathrm{m}$ maps have been smoothed to an angular resolution of $3^{\circ}$ FWHM.}
  \label{fig:hmhd}
\end{figure*}

Since we do not yet have a detailed physical model for the signal,
$\a_{\mathrm{static}}$ must be fitted freely pixel-by-pixel in
solar-centric coordinates for each frequency band. This, however,
introduces a perfect degeneracy between the zero-level of
$\a_{\mathrm{static}}$ and the CIB monopole for the affected channels.
If one adds an arbitrary offset to the map in the top panel in
Fig.~\ref{fig:sidelobe}, and subtracts exactly the same value from the
CIB monopole at the same channel, the total goodness-of-fit at that
channel is unchanged.

As a temporary solution to this problem, we opted in the current data
processing to set the zero-level of the $\a_{\mathrm{static}}$ map at
each channel to the lowest possible value that still results in a
positive signal within instrumental noise fluctuations. The motivation
for this is simplicity of interpretation. Whether the excess signal is
due to zodiacal light emission or sidelobes, it should either way be
positive, and setting it to the lowest possible value ensures that the
final CIB constraints translate into strict upper limits. This is
illustrated in the bottom panel of Fig.~\ref{fig:sidelobe}. In this
case, we have added an extra offset of 0.8\,$\mathrm{MJy\,sr^{-1}}$ to
$\a_{\mathrm{static}}$ at $25\,\mu\mathrm{m}$, or
96$\,\mathrm{nW}\,\mathrm{m}^{-2}\,\mathrm{sr}^{-1}$. Both versions of
this correction template will yield an identical overall $\chi^2$
after fitting the monopole $m_{\nu}$, and both appear as physically
plausible at a purely visual level. However, the bottom case will
yield a CIB monopole that is lower by
0.8\,$\mathrm{MJy\,sr^{-1}}$. The reason for considering precisely
this value here is elaborated on in
Sect.~\ref{sec:results}. Conversely, it is not possible to add a
negative offset of $-0.8$\,$\mathrm{MJy\,sr^{-1}}$, since the template
will then become significantly negative beyond what is allowed by
instrumental noise.

At the same time, the ``lowest possible zero-level'' is not a uniqely
defined quantity, but is rather itself uncertain. Indeed, as described
by \citet{CG02_01}, the specific numerical values adopted for the
absolute zero level, $\a_{\mathrm{static,min}}$, in the final
\cosmoglobe\ DR2 processing were set slightly positive in order to
ensure that the Gibbs sampler explored a physically allowed parameter
region. The difference between these values and the absolute lowest
allowed values were then estimated by convolving the static templates
with a series of different smoothing kernels. The uncertainties
resulting from these calculations are reproduced in column (f) in
Table~\ref{tab:CIB_monopole}, and we add these in quadrature to the
total CIB monopole uncertainty.

In principle, we model the excess static radiation in terms of a free
amplitude, $\a_{\mathrm{static}}$, for each pixel in solar-centric
coordinate map. This map is fitted independently at each wavelength
band between 4.9 and 60$\,\mu\mathrm{m}$, but not in the
others. However, in practice this component is only fitted freely
pixel-by-pixel in the penultimate analysis, as discussed by
\citet{CG02_01}, due to strong degeneracies with the solar components,
and in particular the interplanetary dust cloud. If one lets both the
static component and the cloud parameters to be fitted entirely
freely, the resulting Markov chain has an extremely long correlation
length, and it becomes difficult to assess convergence. To solve this
problem, we therefore only fit the static component pixel-by-pixel in
a preliminary run, and then project out the modes that are orthogonal
to the zodiacal light signal by linear regression, as well as
determine its lowest possible zero-level. This results in the template
seen in the top panel of Fig.~\ref{fig:sidelobe} for the 25$\,\mu$m
channel. In the final production analysis, this component is then kept
fixed; for full discussion about this procedure, see \citet{CG02_01}.

As illustrated in Fig.~\ref{fig:sidelobe}, this template is a high
signal-to-noise ratio contribution, and its inclusion does not
significantly increase the overall noise level of the main
higher-level products. However, since the zero-level of this signal is
unknown -- up to the requirement that it must be positive -- all CIB
monopole results derived for the wavelength range between 4.9 and
60$\,\mu\mathrm{m}$, for which this correction is applied, are
strictly upper limits.

\subsection{Half-mission CIB residual maps}

By applying the algorithm described in Sect.~\ref{sec:algorithm} to the data
summarized in Sect.~\ref{sec:datasummary}, we obtain samples from a full joint
posterior distribution. However, the primary focus of this paper is the CIB
monopole specifically, which, by inspection of
Eqs.~\eqref{eq:model}--\eqref{eq:skymodel}, is actually not explicitly included
in the parametric model at all. Rather, these parameters are only implicitly
included in the general monopole parameter, $m_{\mathrm{\nu}}$, which includes
all contributions to the monopole, of which the CIB is only one. The
appropriate tracer for the CIB monopole in our framework is therefore the
following residual map,
\begin{equation}
\r_{\nu} = \dv_{\nu} - \left(\s_{\mathrm{tot},\nu} - m_{\nu}\right).
\end{equation}
Ideally, if the assumed parametric model were perfect, this map should
consist only of a monopole and instrumental noise.

By computing separate
frequency maps for the first and second half of the mission, we obtain
two different estimates of the true sky at each frequency. We
refer to the first half-mission as HM1 and the second as HM2. We then
define the half-mission half-sum (HMHS) and half-mission
half-difference (HMHD) maps as follows,
\begin{align}
\r_{\nu}^{\mathrm{HMHS}} &= (\r_{\nu}^{\mathrm{HM1}} + \r_{\nu}^{\mathrm{HM2}})/2\\
\r_{\nu}^{\mathrm{HMHD}} &= (\r_{\nu}^{\mathrm{HM1}} -
\r_{\nu}^{\mathrm{HM2}})/2.
\end{align}
Since the true sky signal should be the same in both maps for an ideal
instrument, $\r_{\nu}^{\mathrm{HMHS}}$ provides an estimate of the
true sky, while $\r_{\nu}^{\mathrm{HMHD}}$ provides a combined
estimate of both instrumental noise and residual systematics. Most
importantly for the current analysis, the HMHD residual map includes a
contribution from seasonal modeling errors in the zodiacal light
model. At the same time, it is important to note that both HM1 and HM2
are processed simultaneously in the \cosmoglobe\ DR2 processing
\citep{CG02_01}, and important parameters such as the emissivity and
albedo of zodiacal light parameters are shared between the two
half-missions, and uncertainties in these are not traced by the HMHD
map. In general, uncertainties in parameters that are included in the
parametric model are described by the Markov chain variations, while
seasonally varying model errors are described by the HMHD map. For
full error propagation, both terms must be included.

Figures~\ref{fig:hmhs} and \ref{fig:hmhd} show the residual HMHS and
HMHD maps, respectively, for one single Gibbs sample. The masks
adopted for these figures are defined by two criteria: First, any
included pixel must be observed by both HM1 and HM2. Second, pixels
with a total Galactic foreground contribution larger than a
channel-specific threshold are excluded. We note that both the panel
layout and color ranges are identical between the HMHS and HMHD
figures, and by switching between the two, one can identify
the main features by eye. The bottom two panels have been smoothed to
an angular resolution of $3^{\circ}$ FHWM to suppress instrumental noise.

Two key features are required in order for these data to support a
robust CIB detection, namely 1) a clearly larger positive signal in
the HMHS map than in the HMHD map, and 2) that the HMHS signal appears
statistically isotropic. At the qualitative level of a visual
inspection of Figs.~\ref{fig:hmhs} and \ref{fig:hmhd}, both of these
points appear to hold true for the 1.25, 2.2, 140, and
240\,$\mu\mathrm{m}$ channels. At 3.5$\mu\mathrm{m}$, there are signs
of zodiacal light over-subtraction in the Ecliptic plane, while at
4.9, 12 and 60\,$\mu\mathrm{m}$ the amplitude of the HMHD map is as
strong as in the HMHS map. At 25\,$\mu\mathrm{m}$, there is a large
excess in HMHS, as required, but there is also strong evidence of
zodiacal light and other residuals. At 100\,$\mu\mathrm{m}$, there is
a clear difference between the HMHS and HMHD maps, but there is also
clear evidence of residual Galactic emission. However, even at the
cursory level of such a visual inspection, there appears to be
significant evidence of true CIB monopole signal present in the
\cosmoglobe\ DR2 residual maps.

\subsection{Confidence masks and quality assessment}
\label{sec:masks}

As seen visually in Figs.~\ref{fig:hmhs} and \ref{fig:hmhd}, the
signal-to-noise ratio with respect to pure instrumental noise alone is
massive for all DIRBE channels, and the total uncertainty budget will
ultimately be dominated by astrophysical confusion and instrumental
systematics. With these observations in mind, we define a conservative
set of monopole confidence masks that isolate only the cleanest parts
of the sky through four criteria.

\begin{figure*}
  \centering
  \includegraphics[width=0.376\linewidth]{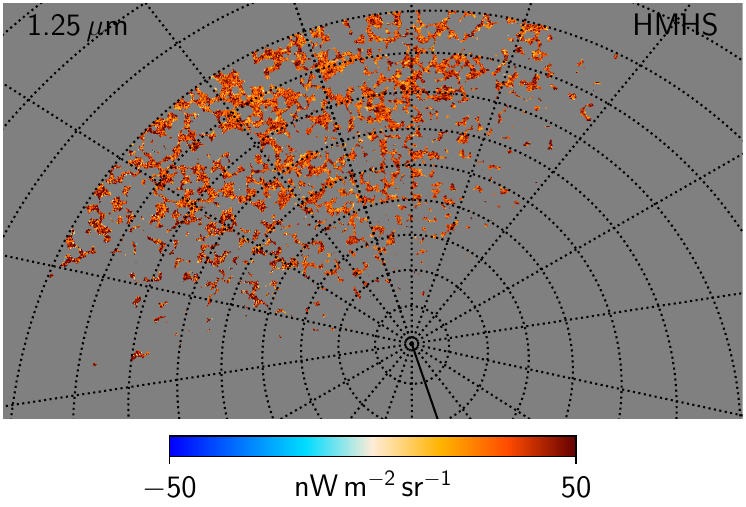}\hspace*{5mm}
  \includegraphics[width=0.376\linewidth]{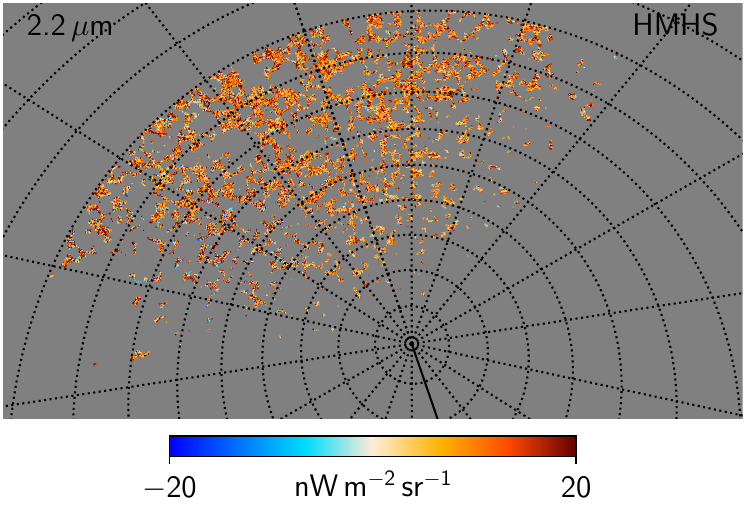}\\
  \includegraphics[width=0.376\linewidth]{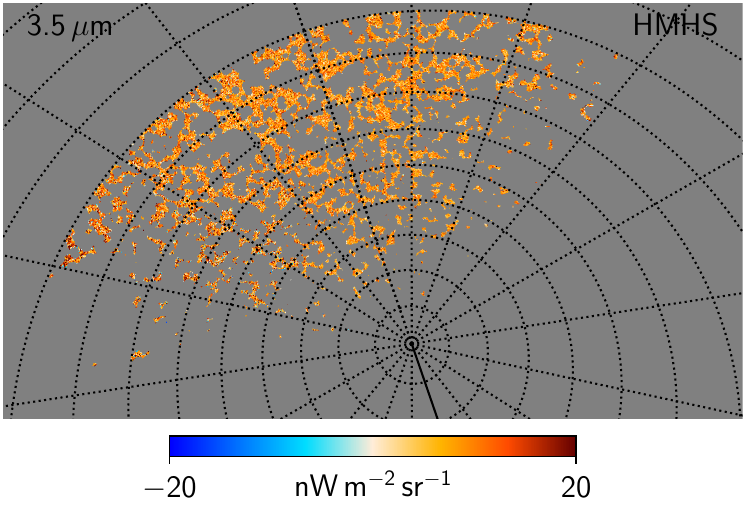}\hspace*{5mm}
  \includegraphics[width=0.376\linewidth]{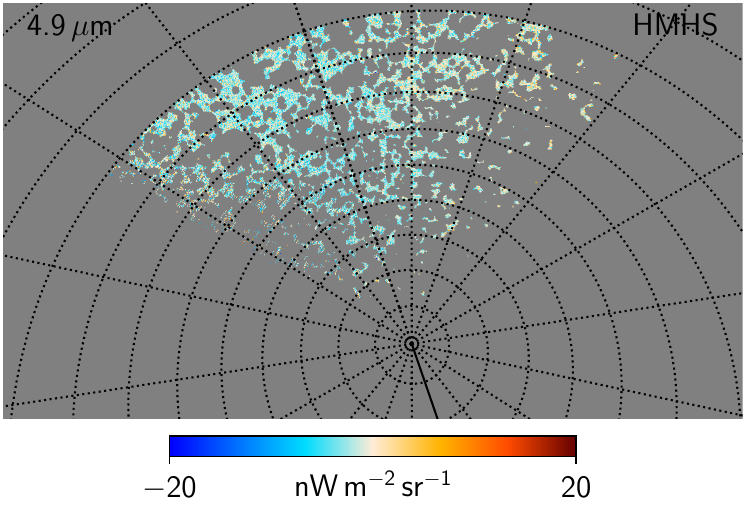}\\
  \includegraphics[width=0.376\linewidth]{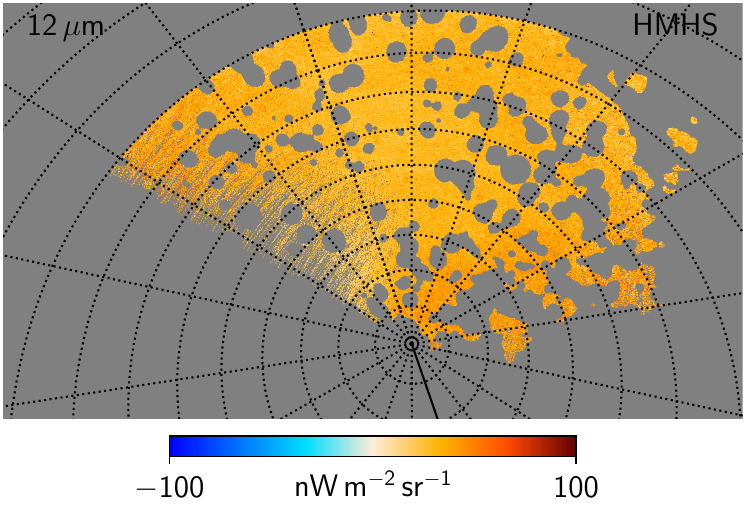}\hspace*{5mm}
  \includegraphics[width=0.376\linewidth]{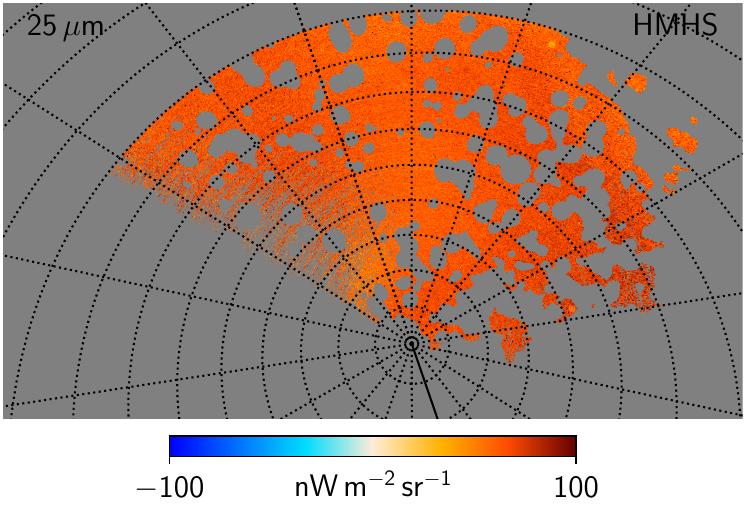}\\
  \includegraphics[width=0.376\linewidth]{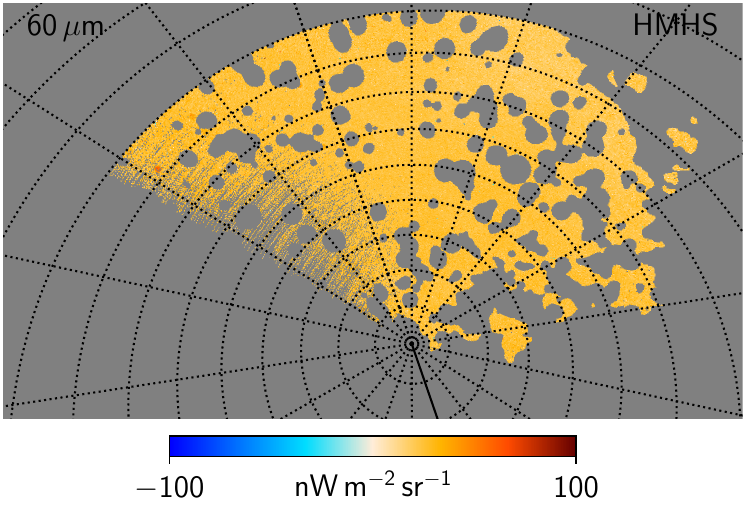}\hspace*{5mm}
  \includegraphics[width=0.376\linewidth]{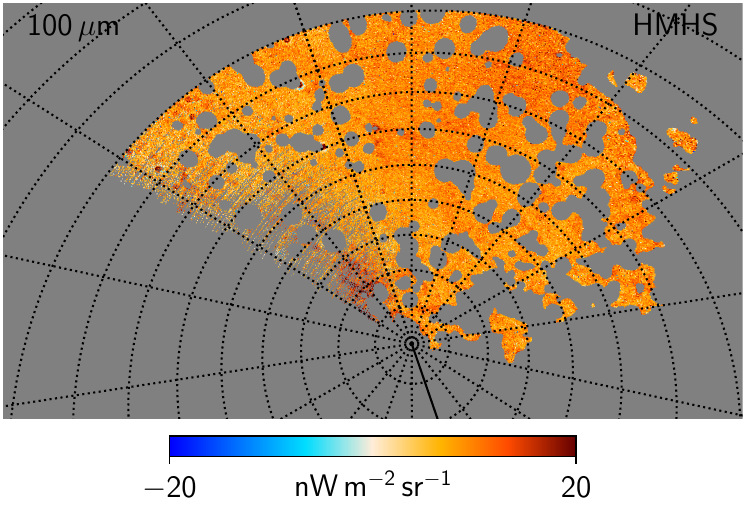}\\
  \includegraphics[width=0.376\linewidth]{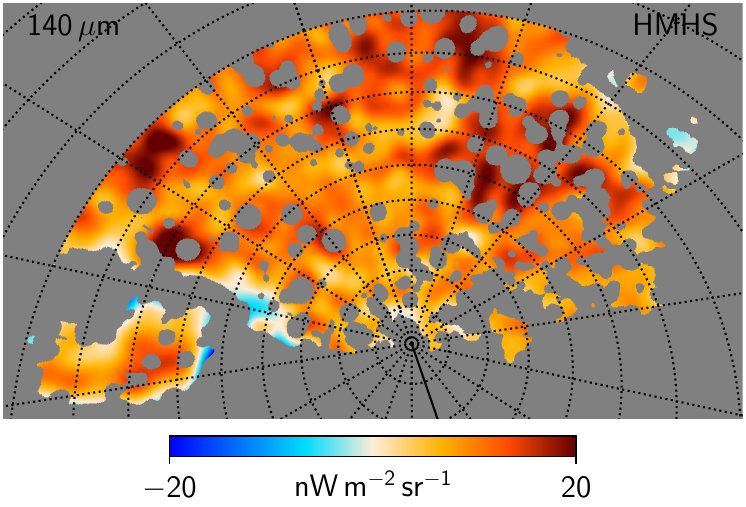}\hspace*{5mm}
  \includegraphics[width=0.376\linewidth]{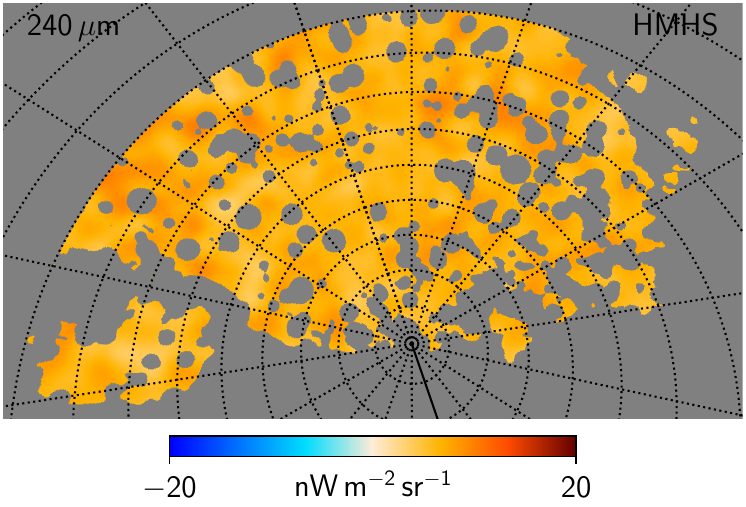}
  \caption{Half-mission half-sum maps, $(\m_{\mathrm{HM1}}+\m_{\mathrm{HM2}})/2$ for each DIRBE frequency channel, zoomed in around the North Ecliptic Pole. Gray pixels indicate the conservative masks used for estimating the monopole. The graticule is centered on the NEP, and the spacing is $5^{\circ}$. The 140 and 240\,$\mu\mathrm{m}$ maps have been smoothed to an angular resolution of $3^{\circ}$ FWHM.} 
  \label{fig:hmhs_zoom}
\end{figure*}

\begin{figure*}
  \centering
  \includegraphics[width=0.376\linewidth]{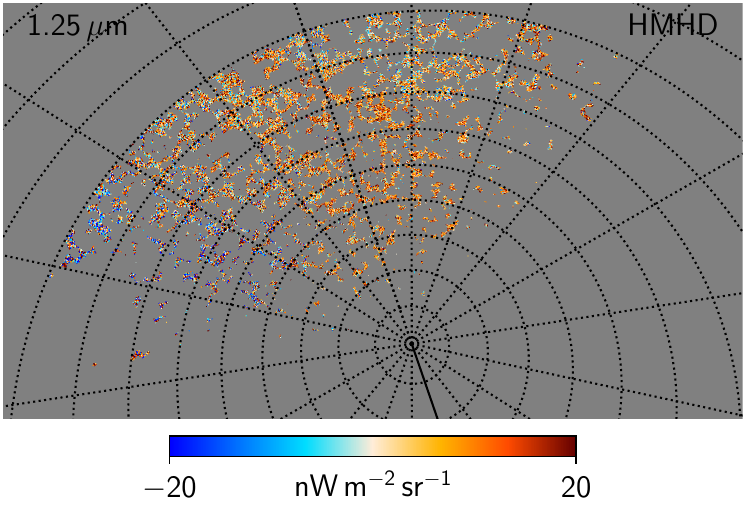}\hspace*{5mm}
  \includegraphics[width=0.376\linewidth]{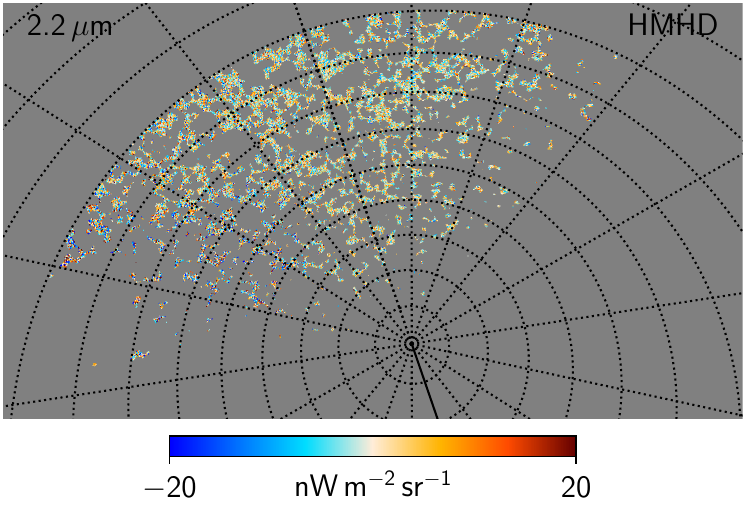}\\
  \includegraphics[width=0.376\linewidth]{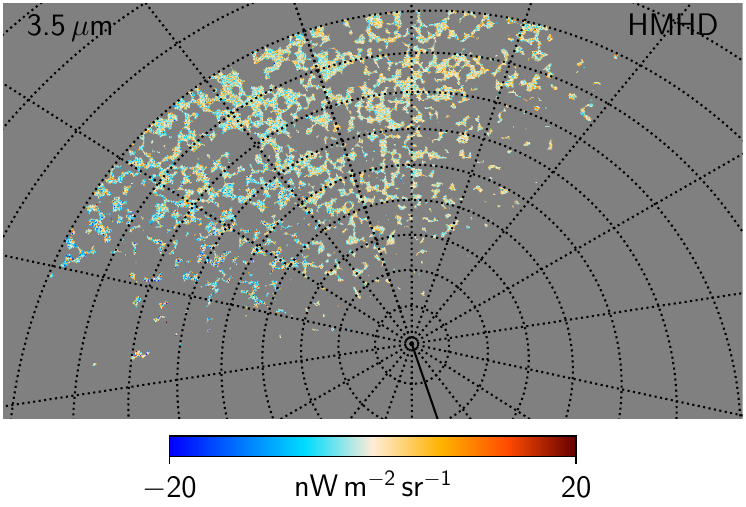}\hspace*{5mm}
  \includegraphics[width=0.376\linewidth]{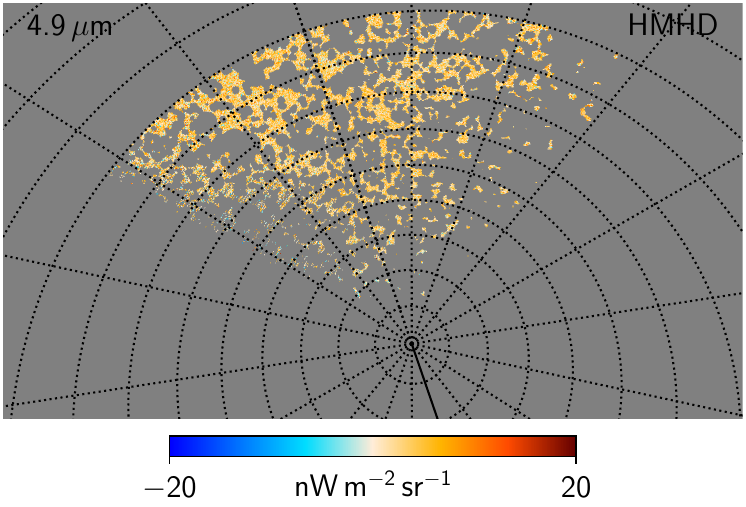}\\
  \includegraphics[width=0.376\linewidth]{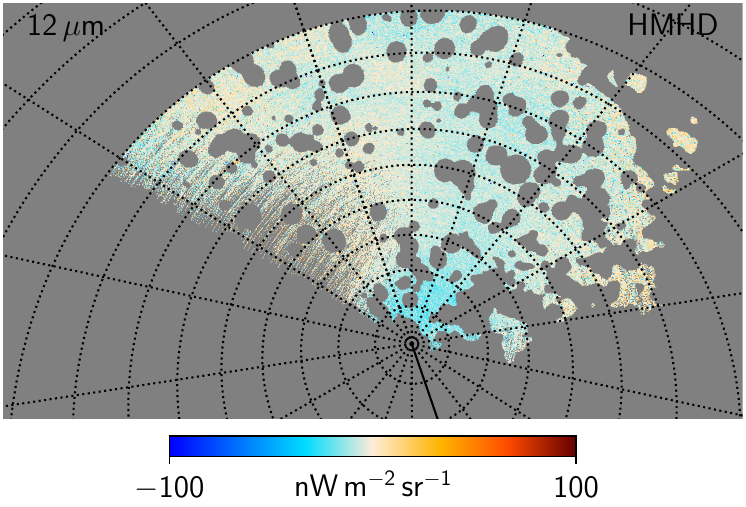}\hspace*{5mm}
  \includegraphics[width=0.376\linewidth]{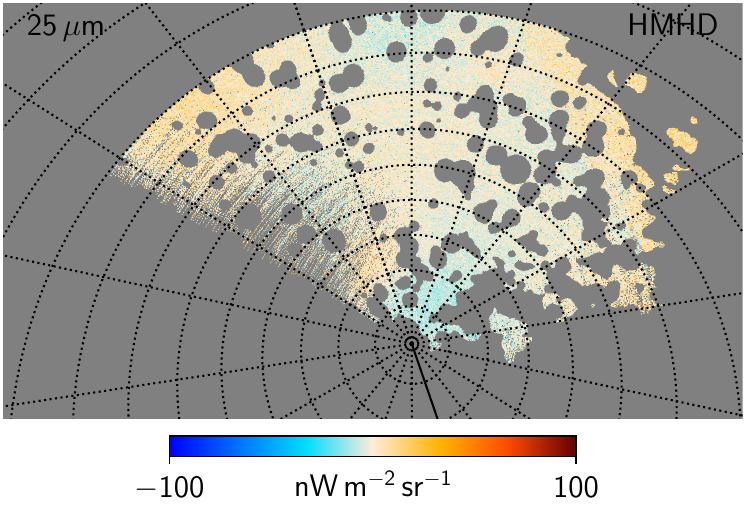}\\
  \includegraphics[width=0.376\linewidth]{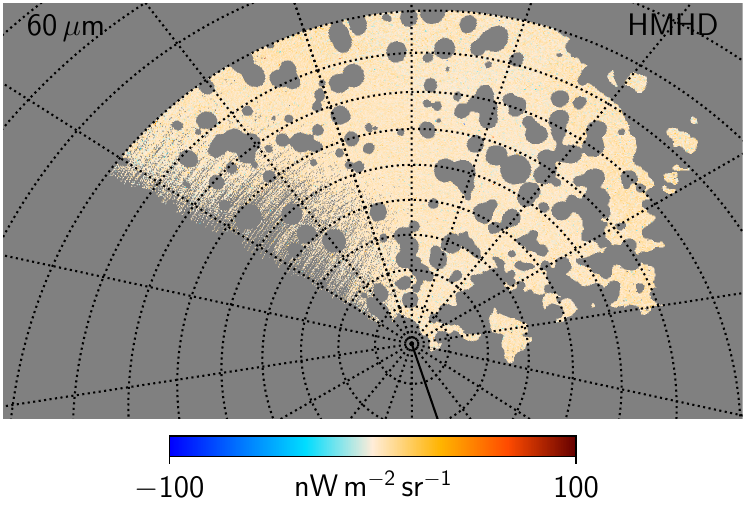}\hspace*{5mm}
  \includegraphics[width=0.376\linewidth]{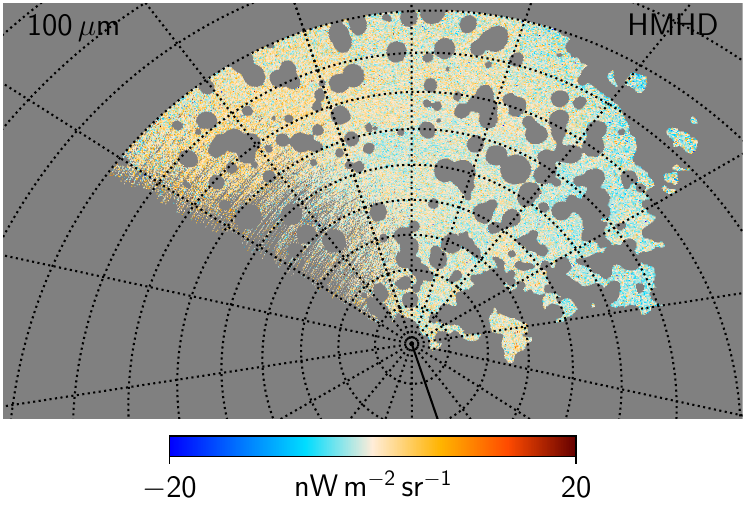}\\
  \includegraphics[width=0.376\linewidth]{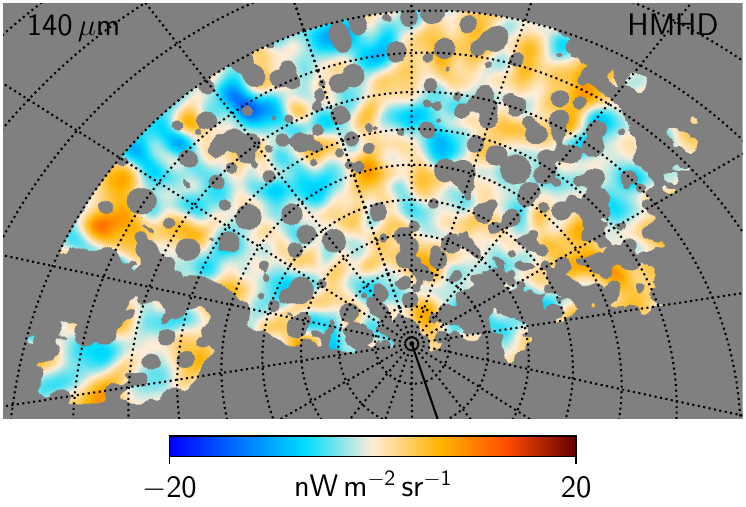}\hspace*{5mm}
  \includegraphics[width=0.376\linewidth]{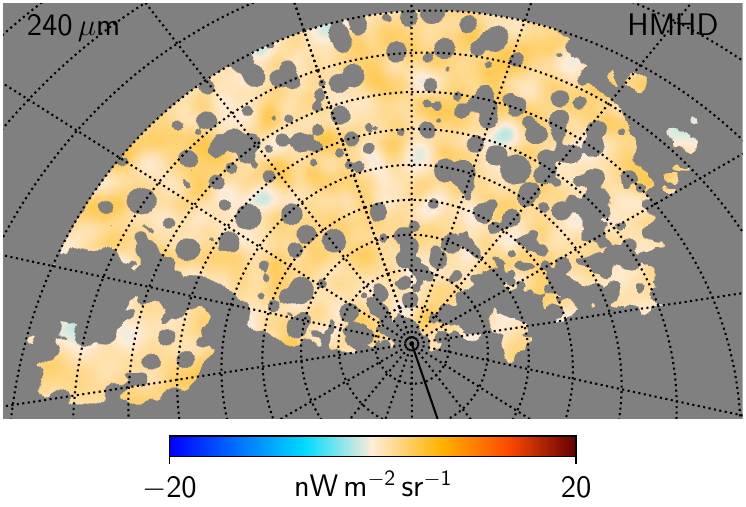}
  \caption{Half-mission half-difference maps, $(\m_{\mathrm{HM1}}-\m_{\mathrm{HM2}})/2$ for each DIRBE frequency channel, zoomed in around the North Ecliptic Pole. Gray pixels indicate the conservative masks used for estimating the monopole. The graticule is centered on the NEP, and the grid spacing is $5^{\circ}$. The 140 and 240\,$\mu\mathrm{m}$ maps have been smoothed to an angular resolution of $3^{\circ}$ FWHM.}
  \label{fig:hmhd_zoom}
\end{figure*}

\begin{figure*}
  \centering
  \includegraphics[width=0.376\linewidth]{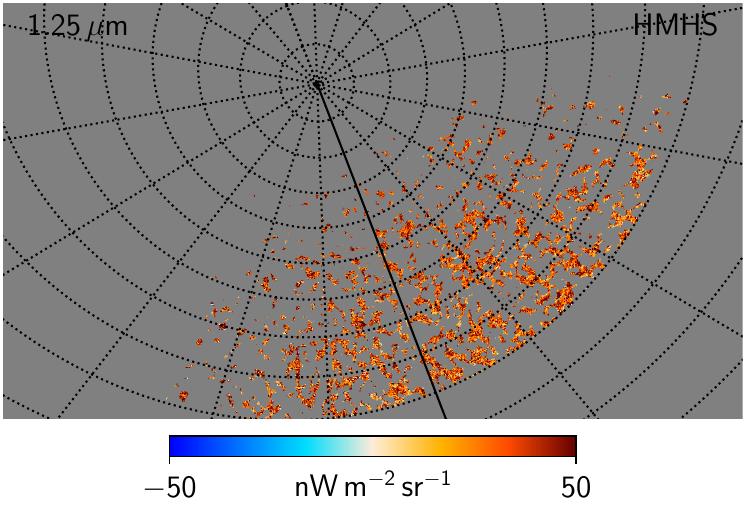}\hspace*{5mm}
  \includegraphics[width=0.376\linewidth]{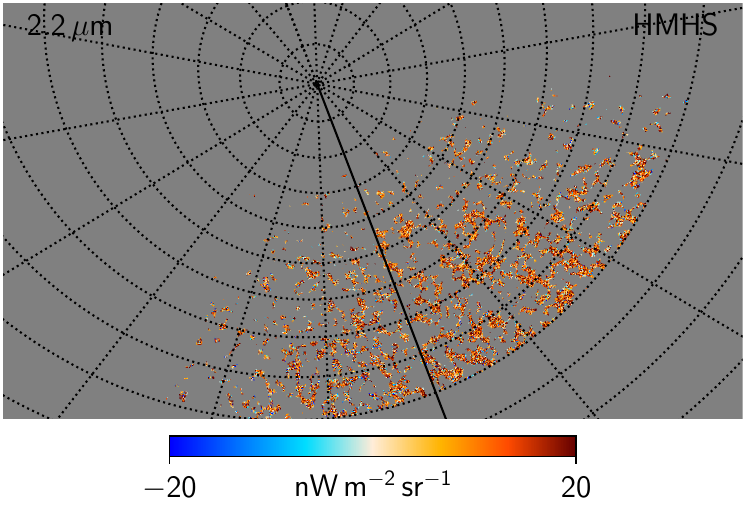}\\
  \includegraphics[width=0.376\linewidth]{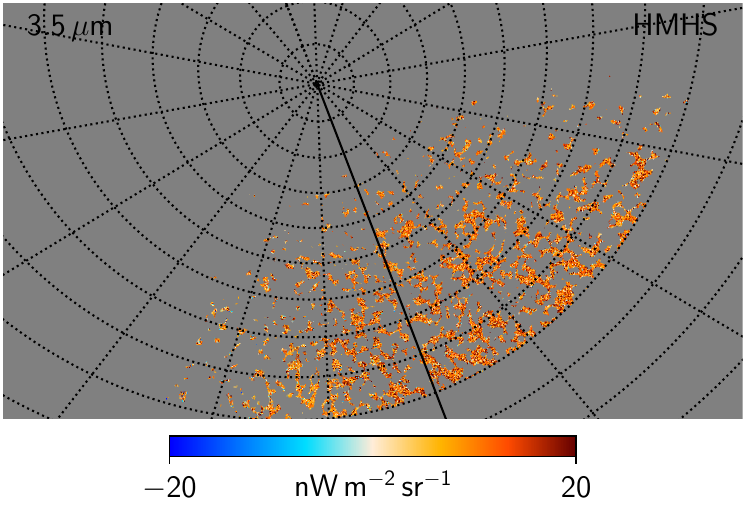}\hspace*{5mm}
  \includegraphics[width=0.376\linewidth]{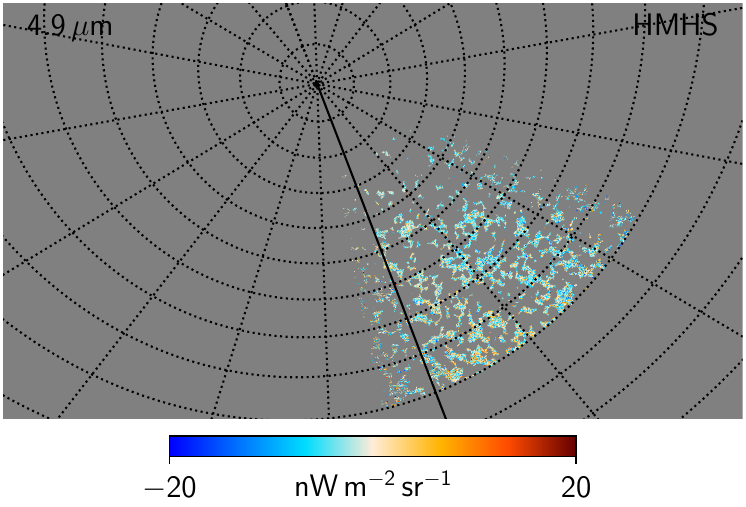}\\
  \includegraphics[width=0.376\linewidth]{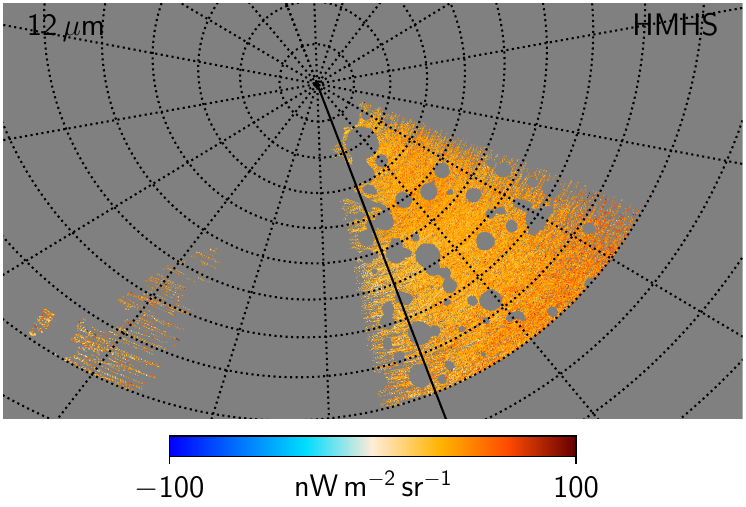}\hspace*{5mm}
  \includegraphics[width=0.376\linewidth]{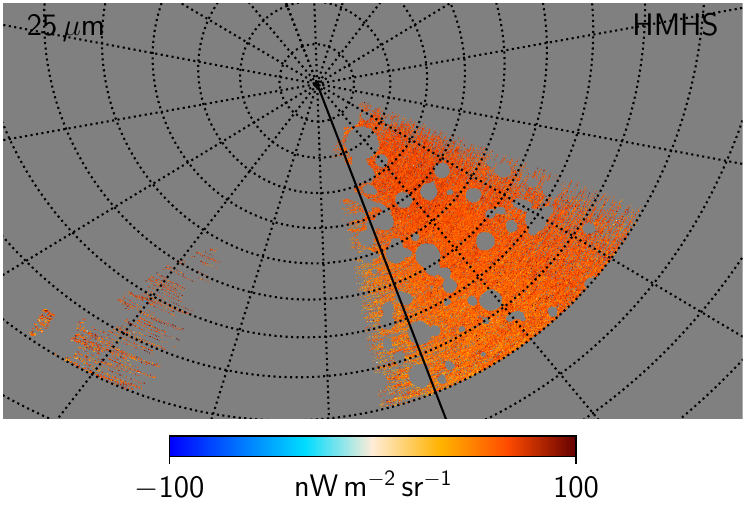}\\
  \includegraphics[width=0.376\linewidth]{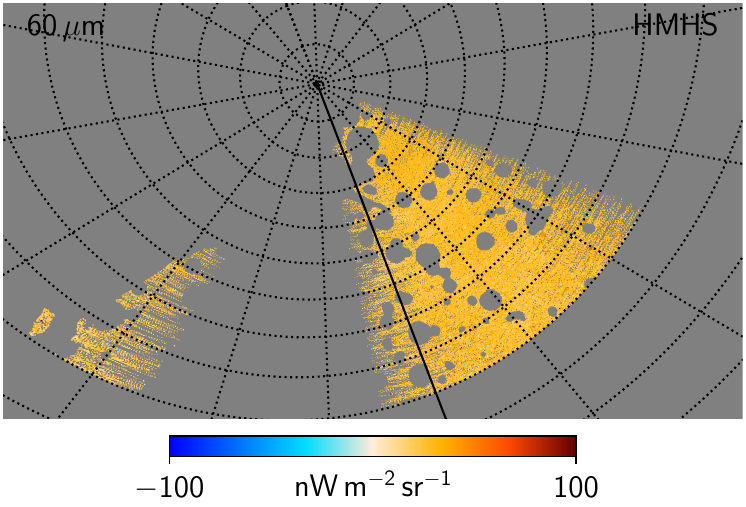}\hspace*{5mm}
  \includegraphics[width=0.376\linewidth]{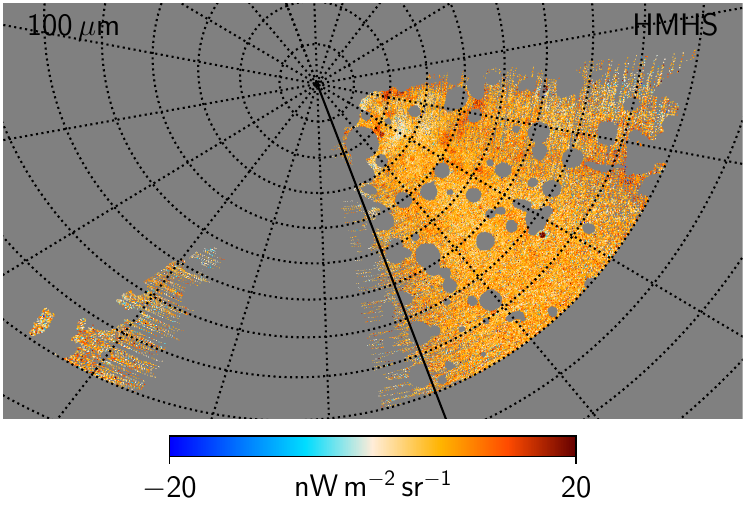}\\
  \includegraphics[width=0.376\linewidth]{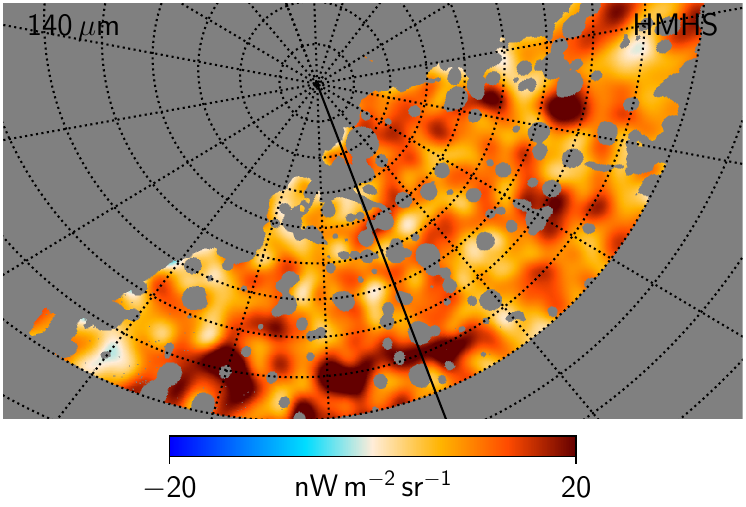}\hspace*{5mm}
  \includegraphics[width=0.376\linewidth]{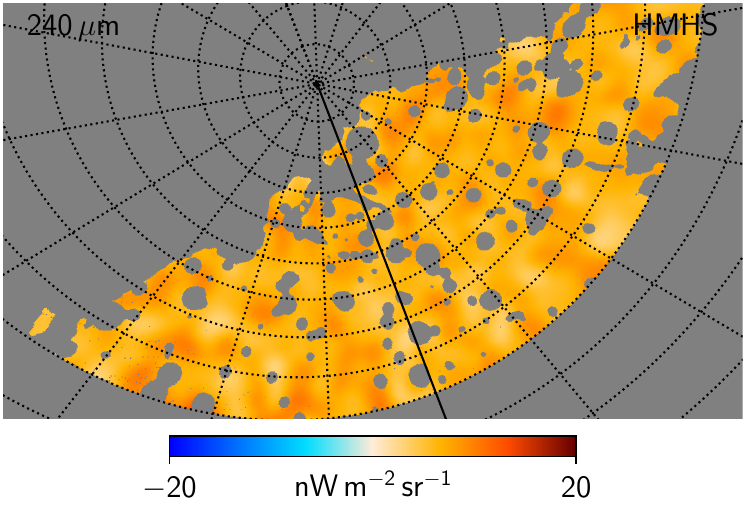}
  \caption{Same as Fig.~\ref{fig:hmhs_zoom}, but centered on the South Ecliptic Pole.}
  \label{fig:hmhs_zoom_south}
\end{figure*}

\begin{figure*}
  \centering
  \includegraphics[width=0.376\linewidth]{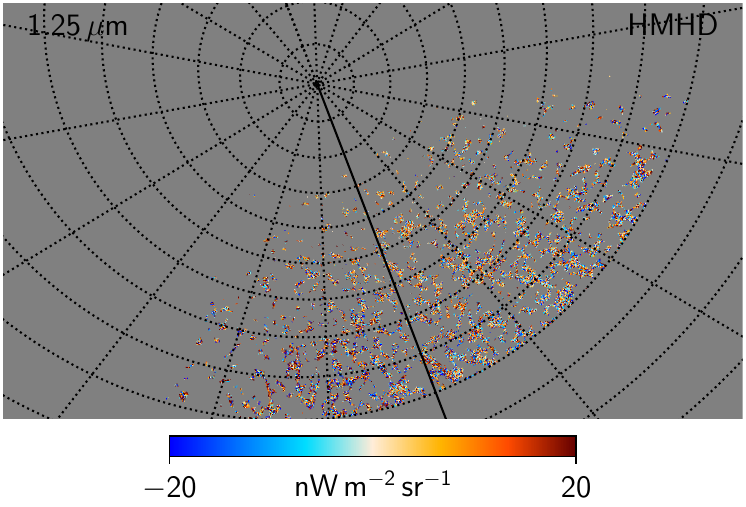}\hspace*{5mm}
  \includegraphics[width=0.376\linewidth]{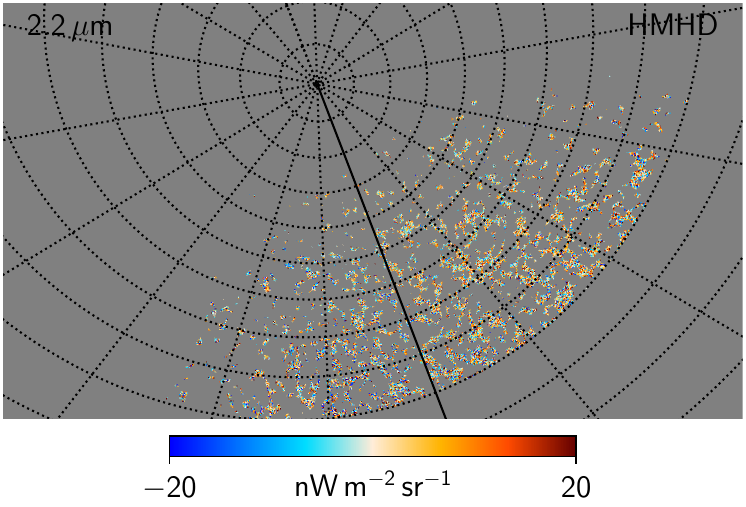}\\
  \includegraphics[width=0.376\linewidth]{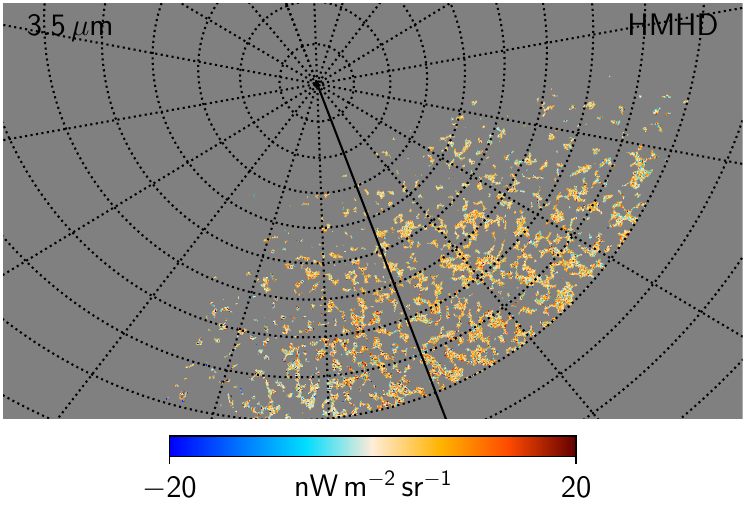}\hspace*{5mm}
  \includegraphics[width=0.376\linewidth]{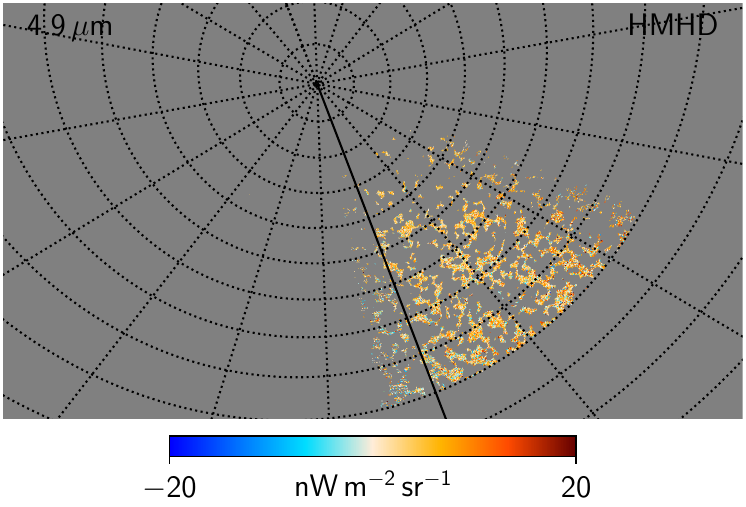}\\
  \includegraphics[width=0.376\linewidth]{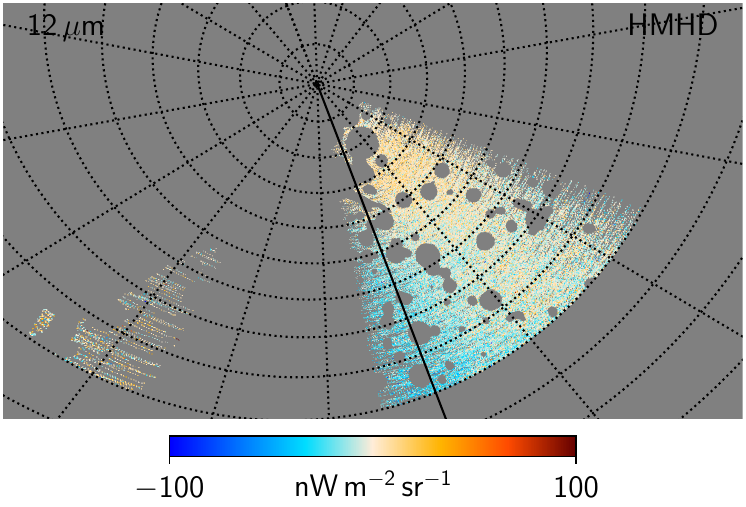}\hspace*{5mm}
  \includegraphics[width=0.376\linewidth]{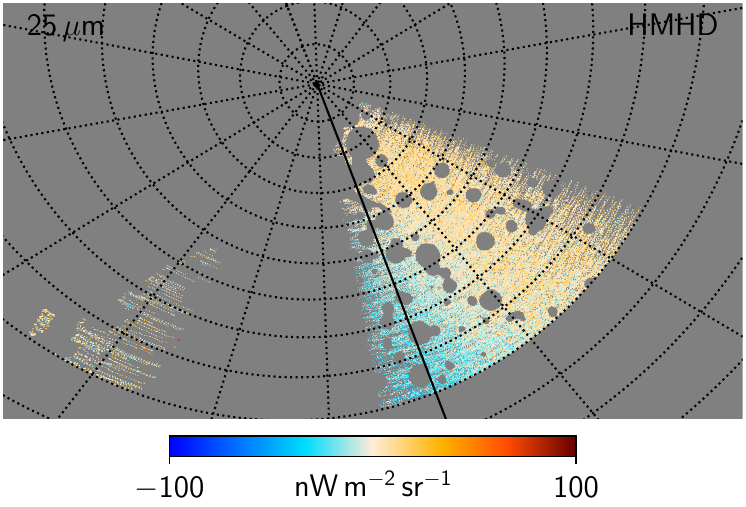}\\
  \includegraphics[width=0.376\linewidth]{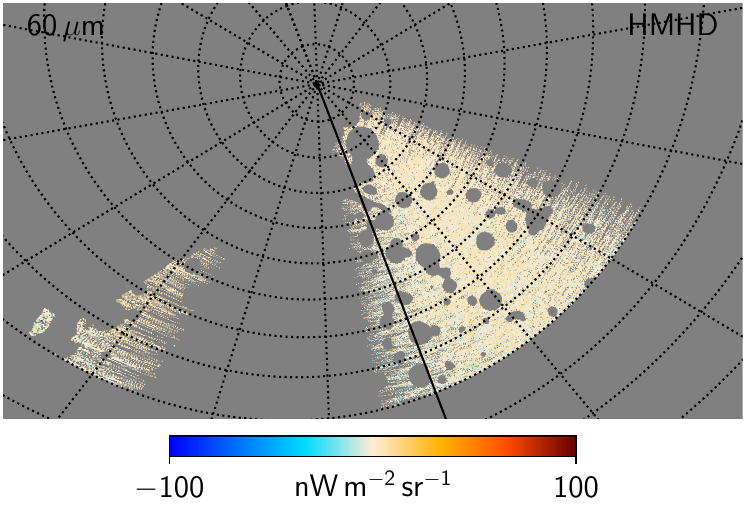}\hspace*{5mm}
  \includegraphics[width=0.376\linewidth]{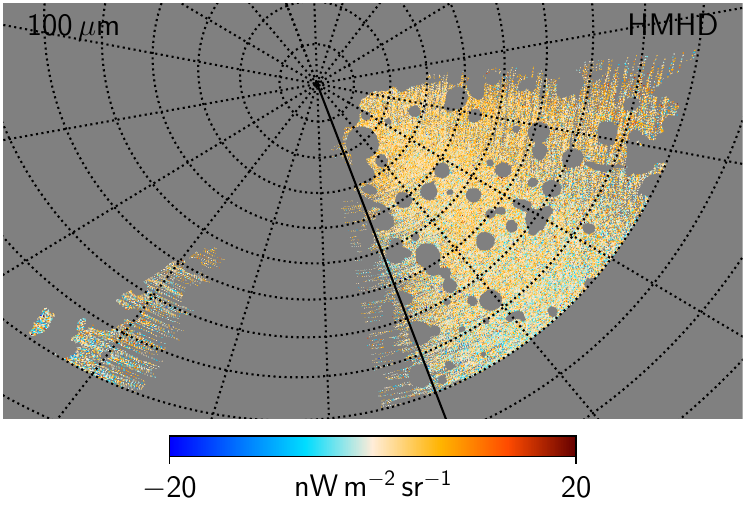}\\
  \includegraphics[width=0.376\linewidth]{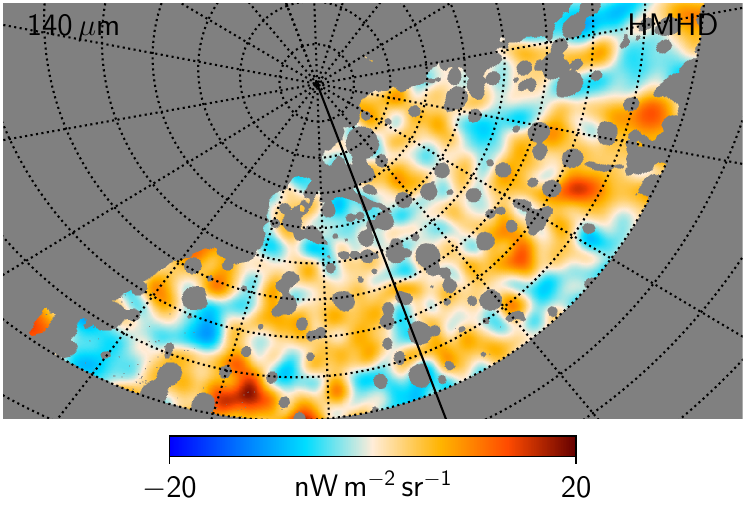}\hspace*{5mm}
  \includegraphics[width=0.376\linewidth]{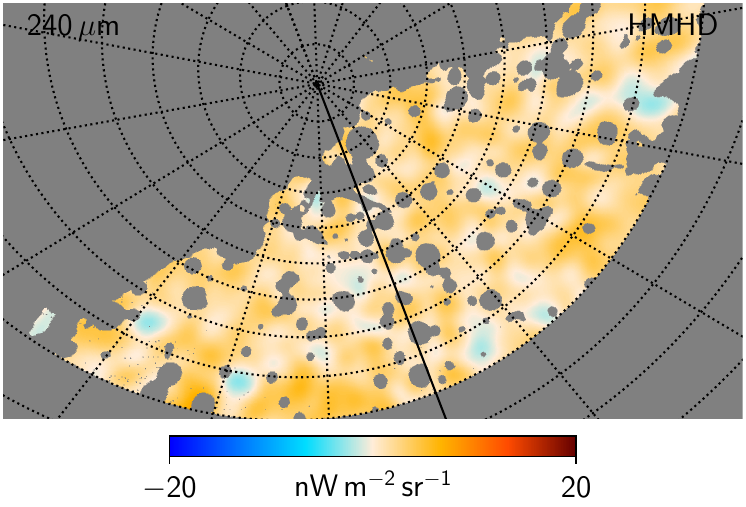}
  \caption{Same as Fig.~\ref{fig:hmhd_zoom}, but centered on the South Ecliptic Pole.}
  \label{fig:hmhd_zoom_south}
\end{figure*}

First, following the above prescription, we require that any
accepted pixel must be observed by both HM1 and HM2. This is a strict
requirement in order to be able to use the HMHD maps directly for
error propagation over the same sky fraction as used for estimating
the signal level itself. 

\begin{figure}
  \centering
  \includegraphics[width=0.98\linewidth]{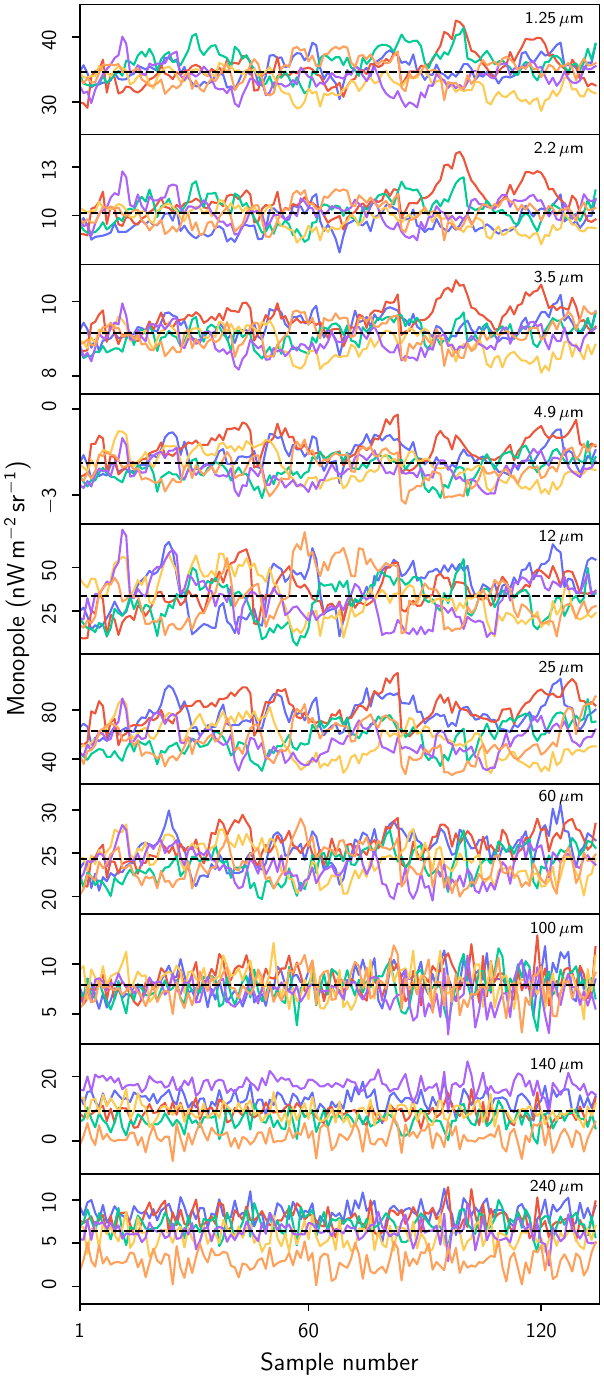}
  \caption{CIB monopole estimates as a function of Gibbs sample iteration for each DIRBE channel. Each color shows one Markov chain, and the horizontal dashed line shows the final \Cosmoglobe\ DR2 posterior mean values as tabulated in Table~\ref{tab:CIB_monopole}. The first 20 samples from each chain have already been removed as burn-in.  }
  \label{fig:traceplot}
\end{figure}

Second, it is evident from Fig.~\ref{fig:hmhd} that the Ecliptic
plane is particularly susceptible to zodiacal light residuals. We
therefore exclude all pixels with an absolute Ecliptic latitude below
$|b|<45^{\circ}$ from the analysis; this cut excludes 71\% of the
sky. We have checked that setting the limit at either $|b|<60^{\circ}$
or $75^{\circ}$ gives very similar results, only with slightly larger
Monte Carlo uncertainties.

Third, to remove pixels with obvious modeling failures, we
evaluate the so-called $\chi^2$ map of the form
\begin{equation}
\chi_p^2 = \sum_{\nu} \left(\frac{d_{\nu,p}-s_{\nu,p}^{\mathrm{sky}}}{\sigma_{\nu,p}}\right)^2,
\end{equation}
where $p$ is defined at a HEALPix\footnote{\url{http://healpix.sourceforge.io}} grid of $N_{\mathrm{side}}=512$,
corresponding to $7\arcmin\times 7\arcm$ pixels
\citep{healpix}. Contributions from the \Planck\ frequency bands,
which have four times higher resolution, are co-added into these
coarser pixels, and each of the six \Planck\ channels therefore
contributes with 16 pixels per $\chi^2$ element. We then smooth this
$\chi^2$ map to $1^{\circ}$ FWHM, and remove any pixel with a value
larger than 200, corresponding roughly to a reduced $\chi^2$ of 1.9;
this cut excludes 14\,\% of the sky. We have checked that increasing the
threshold by a factor of two or five does not significantly affect the
results.

Finally, we remove any pixels with a large absolute Galactic
foreground contribution. For channels between 1.25 and
4.9\,$\mu\mathrm{m}$, we exclude any pixels for which the sum of the
bright stars and other compact objects (as defined by
Eq.~\ref{eq:skymodel}) evaluated at 1.25\,$\mu\mathrm{m}$ is brighter
than 20\,$\mathrm{kJy\,sr^{-1}}$, or where the faint starlight template is brighter
than 50\,$\mathrm{kJy\,sr^{-1}}$. Combined, these cuts exclude 88\,\% of the sky. For
the six longer-wavelength bands, we exclude any pixels for which the
sum of the three dust components is larger than 50\,$\mathrm{MJy\,sr^{-1}}$ evaluated
at the pivot frequency of 545\,$\mathrm{MJy\,sr^{-1}}$; this removes 32\,\% of the sky.

Figures~\ref{fig:hmhs_zoom} and \ref{fig:hmhd_zoom} shows zoom-ins
around the North Ecliptic Pole of the same HMHS and HMHD maps plotted
in Figs.~\ref{fig:hmhs} and \ref{fig:hmhd}, but with the new and more
conservative analysis masks applied. Similar zoom-ins of around the South Ecliptic Pole are shown in Figs.~\ref{fig:hmhs_zoom_south} and \ref{fig:hmhd_zoom_south}. Again, when comparing the HMHS
and HMHD maps in these figures, the 1.25, 2.2, 140, and
240$\,\mu\mathrm{m}$ channels all appear to provide a highly
significant detection of an isotropic signal. Indeed, with these more
stringent cuts, even the 3.5 and 100$\,\mu\mathrm{m}$ channels appear
sufficiently clean to justify a direct measurement. 

Finally, we note that the 25\,$\mu\mathrm{m}$ channel appears rather
anomalous in this data set. Specifically, even though it actually
appears rather isotropic by visual inspection, it has a higher
amplitude than either of the two neighboring channels, namely a value
of about 50$\,\mathrm{nW}\,\mathrm{m}^{-2}\,\mathrm{sr}^{-1}$ compared
to less than 30$\,\mathrm{nW}\,\mathrm{m}^{-2}\,\mathrm{sr}^{-1}$ for
the 12 and 60$\,\mu\mathrm{m}$ channels. Such a rapidly changing
monopole spectrum is obviously very difficult to explain in terms of
either astrophysics or a cosmological signal. A far more compelling
explanation is the uncertainty in the zero-level of the static
component shown in Fig.~\ref{fig:sidelobe}. The current results are
derived with the default low zero-level template shown in the top
panel. However, if the true zero-level of the 25$\,\mu$m channel
should happen to be 0.8\,$\mathrm{MJy\,sr^{-1}}$, the excess seen in
Fig.~\ref{fig:hmhs_zoom} would vanish entirely.

\begin{table*}
\newdimen\tblskip \tblskip=5pt
\caption{Summary of CIB monopole constraints and uncertainties. All monopoles and uncertainties are given in units of $\nWmsr$. For channels with a robust monopole detection, the central value corresponds to the posterior mean of all accepted Gibbs samples, and the final DR2 uncertainty is given in column (g); for channels without a robust monopole detection, the upper limit is defined as the sum of the posterior mean value and twice the total uncertainty. }
\label{tab:CIB_monopole}
\vskip -4mm
\footnotesize
\setbox\tablebox=\vbox{
 \newdimen\digitwidth
 \setbox0=\hbox{\rm 0}
 \digitwidth=\wd0
 \catcode`*=\active
 \def*{\kern\digitwidth}
  \newdimen\dpwidth
  \setbox0=\hbox{.}
  \dpwidth=\wd0
  \catcode`!=\active
  \def!{\kern\dpwidth}
  \halign{\hbox to 1.7cm{#\leaderfil}\tabskip 2em&
    \hfil$#$\hfil \tabskip 2em& %
    \hfil$#$\hfil \tabskip 1em& %
    \hfil$#$\hfil \tabskip 1em&
    \hfil$#$\hfil \tabskip 1em& 
    \hfil$#$\hfil \tabskip 1em&
    \hfil$#$\hfil \tabskip 1em&
    \hfil$#$\hfil \tabskip 2em&
    \hfil$#$\hfil \tabskip 1em&%
    \hfil$#$\hfil \tabskip 1em&
    \hfil$#$\hfil \tabskip 2em&
    \hfil$#$\hfil \tabskip 1em& %
    \hfil$#$\hfil \tabskip 0em\cr
\noalign{\doubleline}
\omit&&\multispan6\hfil\sc Monopole uncertainty\hfil&\multispan3\hfil\sc CIB monopole constraint \hfil&\multispan2\hfil\sc Sky fraction (\%) \hfil\cr
\noalign{\vskip -3pt}
\omit& &\multispan6\hrulefill&\multispan3\hrulefill&\multispan2\hrulefill\cr
\noalign{\vskip 3pt} 
\omit\sc $\lambda$ ($\mu\mathrm{m}$)\hfil& U_{B_\nu\rightarrow\nu I_\nu}^{(\mathrm{a})} &\sigma_b^{\mathrm{(b)}} & \sigma_g^{\mathrm{(c)}} & \sigma_\mathrm{MC}^{\mathrm{(d)}} & \sigma_\mathrm{HM}^{\mathrm{(e)}}  & \sigma^{\mathrm{(f)}}_\mathrm{static}  & \sigma_\mathrm{Total}^{\mathrm{(g)}}  & \mathrm{DIRBE}^{\mathrm{(h)}} & \mathrm{Gibbs}^{(\mathrm{i})} & \mathrm{Final\,DR2}^{(\mathrm{j})} & \mathrm{Gibbs}^{(\mathrm{k})} & \mathrm{Final\,DR2}^{(\mathrm{l})} \cr
\noalign{\vskip 3pt\hrule\vskip 5pt}
*1.25 & 2400 & 0.05 & 1.0 & *1.6  & *5.1   & 0!* & *5.5     & <75        & 34\pm10  & *35\pm6 & 18 & 1.9 \cr
*2.2  & 1364 & 0.03 & 0.3 & *0.6  & *1.0   & 0!* & *1.2     & <39        & 10.8\pm1.7  & *10.2\pm1.2 & 18 & 1.9 \cr
*3.5  & *857 & 0.02 & 0.3 & *0.3  & *1.2   & 0!* & *1.3     & <23        & *8.0\pm2.2  & **9.2\pm1.3 & 18 & 1.9 \cr
*4.9  & *612 & 0.01 & 0.1 & *0.4  & *3.4   & 2.4 & *4.2 & <41            & *<13     & **<8    & 18 & 1.3 \cr
*12   & *250 & 0.02 & 1.7 & *8.9  & *5.1   & 1.2 & 10.5 & <468           & *<68     & *<55    & 81 & 4.7 \cr
*25   & *120 & 0.01 & 9.5 & 11.8  & *0.2   & 1.0 & 15.2 & <504           & *<89     & *<93    & 81 & 4.7 \cr
*60   & **50 & 1.34 & 2.6 & *1.5  & *2.3   & 1.8 & *4.4 & <75            & *<35     & *<33    & 48 & 4.8 \cr
100   & **30 & 0.81 & 1.0 & *1.0  & *0.6   & 0!* & *1.7 & <34            & *8.8\pm1.8  & **7.9\pm1.7    & 48 & 5.2 \cr
140   & **21 & 5!** & 1.1 & *3.9  & *0.4   & 0!* & *6.4 & 25.0\pm6.9     & *8\pm6  & **9\pm6     & 52 & 8.8 \cr
240   & **12 & 2!** & 0.7 & *1.5  & *1.5   & 0!* & *3.0 & 13.6\pm2.5     & *6.1\pm2.3  & **6\pm3     & 52 & 8.8 \cr
\noalign{\vskip 5pt\hrule\vskip 5pt}}}
\endPlancktablewide
\tablenote {{\rm a}} Intensity unit conversion factor, $U_{B_\nu\rightarrow\nu I_\nu} = 3000/\lambda[\mu\mathrm{m}]$, in units of $(\mathrm{nW}\,\mathrm{m}^2\,\mathrm{sr}^{-1}) / (\mathrm{MJy}\,\mathrm{sr}^{-1})$.\par
\tablenote {{\rm b}} CIO baseline (or offset) uncertainty as estimated by the DIRBE team; reproduced from Table~1 of \citet{hauser1998}.\par
\tablenote {{\rm c}} CIO gain uncertainty; estimated by multiplying
the gain uncertainty in Table~1 of \citet{hauser1998} with the
\cosmoglobe\ posterior mean values.\par
\tablenote {{\rm d}} Statistical Monte Carlo uncertainty estimated as the standard deviation of all accepted Gibbs samples; accounts for astrophysical and zodiacal light uncertainties.\par
\tablenote {{\rm e}} Systematic monopole uncertainty, defined as the mean absolute difference between individual HM1 and HM2 estimates; accounts for zodiacal light modeling errors and potential instrumental drifts.\par
\tablenote {{\rm f}} Systematic monopole uncertainty from the unknown zero-level of the sidelobe model; see \citet{CG02_01} for details.\par
\tablenote {{\rm g}} Total monopole uncertainty obtained by adding the individual uncertainties in columns (b)--(f) in quadrature.\par
\tablenote {{\rm h}} Official DIRBE monopole constraints reproduced from Table~1 of \citet{hauser1998}.\par
\tablenote {{\rm i}} \cosmoglobe\ DR2 CIB constraint as derived directly from the monopole parameter in the Gibbs chain; see \citet{CG02_01}.\par
\tablenote {{\rm j}} \cosmoglobe\ DR2 CIB constraint as derived with the tuned monopole masks discussed in Sect.~\ref{sec:masks}.\par
\tablenote {{\rm k}} Sky fraction used for sampling $m_{\nu}$ in the main \cosmoglobe\ DR2 analysis; see \citet{CG02_01}.\par
\tablenote {{\rm l}} Sky fraction used for estimating the final posterior mean monopole with the masks defined in Sect.~\ref{sec:masks}.\par
\par
\end{table*}

\begin{figure*}
	\centering
	\includegraphics[width=\textwidth]{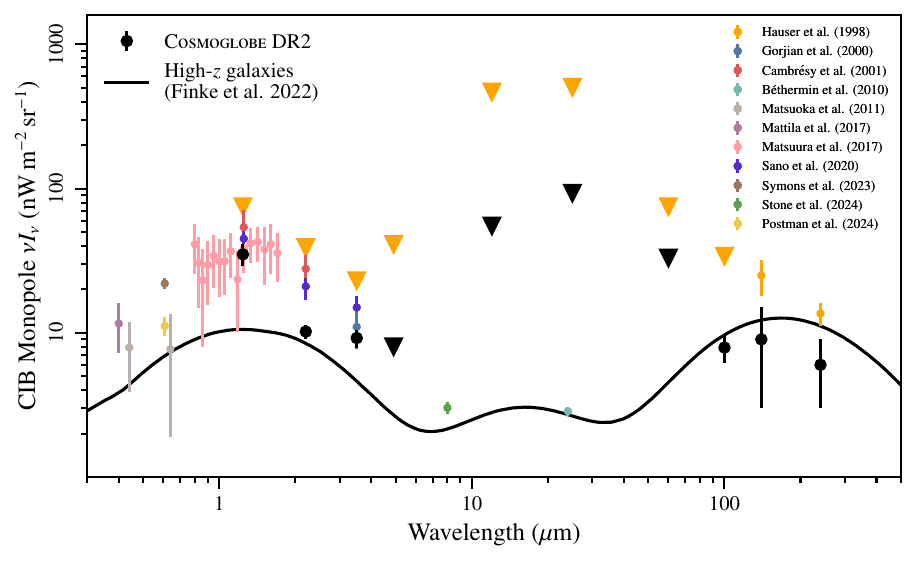}
	\caption{
		Comparison of theoretical and observational constraints on the CIB monopole. Black lines include models of integrated galactic background line \citep{finke2022} and Population III stars \citep{santos:2002}. Black points show the measurements and upper limits from this work. The orange, blue, red, and dark purple points show independent analyses of the legacy DIRBE maps \citep{hauser1998,gorjian:2000,cambresy:2001,sano:2020}, while the cyan and red points show results from \textit{Spitzer} \citep{bethermin:2010} and \textit{JWST} \citep{stone:2024}. Gray points \citep{matsuoka:2011} are from the \textit{Pioneer 10} IPP, light purple \citep{mattila:2017} from ESO VLT, pink \citep{matsuura:2017} from \textit{CIBER}, and brown \citep{symons:2023} and yellow \citep{postman:2024} from New Horizons.
		All upper limits are denoted by downward-facing triangles at the 95\,\% upper limit as calculated in Table \ref{tab:CIB_monopole}, while all error bars are 68\,\% confidence intervals.
	}
	\label{fig:CIB_monopoles}
\end{figure*}

With this visually obvious observation in mind, it is worth
re-emphasizing that the same argument applies also to the other
channels for which the static template is applied, namely
4.9--60$\,\mu\mathrm{m}$. By increasing the zero-level of the
respective static template, the monopole seen in
Fig.~\ref{fig:hmhs_zoom} can be decreased all the way to
zero. However, since the static template has to be positive, the monopole
cannot be increased significantly compared to what is seen in
Fig.~\ref{fig:hmhs_zoom}. These maps do therefore impose strong upper
limits on the CIB monopole in this wavelength range, despite the
presence of the static component.

\section{Results}
\label{sec:results}

\subsection{Burn-in and Monte Carlo convergence}

As described in Sect.~\ref{sec:algorithm}, the \cosmoglobe\ algorithm is a
Gibbs sampler, and is as such subject to Monte Carlo burn-in and
convergence. It has to run for long enough to reach a stationary
state, and once it does that, it has to explore the full joint
posterior distribution for sufficiently long to ensure that the
posterior standard deviation is well measured.

Figure~\ref{fig:traceplot} shows the mean monopole as a function of
Gibbs iteration for each DIRBE channel as evaluated over the
confidence masks defined in Sect.~\ref{sec:masks}, after removing the
first 20 samples in each chain as suggested by \citet{CG02_02}. The
colored lines show results from the six independent Gibbs
chains. Based on this plot, we do not see any significant evidence for
remaining burn-in, as the chains appear stationary from the very
beginning.

As far as Monte Carlo convergence goes, we see that the remaining
samples appear to scatter with a relatively short correlation length
for all channels at wavelengths shorter than
140$\,\mu\mathrm{m}$. However, at the two longest wavelengths the
correlation length is very long, and each chain appears to largely explore
its own local minimum. As discussed by \citet{CG02_02}, these channels
also have a low signal-to-noise ratio with respect to zodiacal light,
and there is therefore a strong degeneracy between the band monopole
and the zodiacal light emissivity that takes a long time to explore
with Gibbs sampling. At the same time, the same analysis also shows
that the combination of the six chains does cover the physically
plausible range for the zodiacal light emissivities, and the range
covered by the monopole chains seen in Fig.~\ref{fig:traceplot}
therefore also provide a useful estimate of the corresponding
physically meaningful monopole uncertainty, despite the long
correlation length. Indeed, in a future update of this analysis it
might be prudent to impose a physically motivated prior on the
zodiacal light emissivity by extrapolating from the
12--60\,$\mu\mathrm{m}$ channels, and this will then significantly
decrease the Monte Carlo uncertainties seen in this figure. For now,
however, we prefer the conservative approach, and do not impose any
priors on the zodiacal light amplitude, at the cost of increased
monopole uncertainties in the far-infrared regime.

\subsection{Monopole constraints}

We are now finally ready to present one of the main results from the
\cosmoglobe\ DR2 analysis, namely updated constraints on the CIB
monopole spectrum from DIRBE. Based on the above discussions, we
provide point measurements at 1.25, 2.2, 3.5, 100, 140, and
240$\,\mu\mathrm{m}$, while for the remaining channels we provide only
upper limits.

For channels with a positive detection, the central value is simply
taken as the posterior mean averaged over all accepted Gibbs
samples. In contrast, the corresponding uncertainty is significantly
more complicated, and includes five different terms added in
quadrature. The first contribution is simply the posterior RMS as
evaluated from the Gibbs samples; this quantifies statistical
uncertainties that are directly described by the parametric model in
Eqs.~\eqref{eq:model}--\eqref{eq:skymodel}. The most important
examples of such are zodiacal light and astrophysical foreground
variations. Second, we include a contribution defined by the absolute
value of the monopole of the HMHD map. This measures modeling errors
that are not captured within the model itself, but has a seasonal
variation; a typical example of such is zodiacal light mismodeling
errors that leads to different signatures in the first and second half
of the DIRBE survey. Third, we include a term that describes residual
uncertainties in the zero-level of the static component. As discussed
above, the zero-level of these templates have been set as low as
possible without introducing large negative regions. However, this
value itself is not unique, but rather depends for instance on the
intrinsic noise level of the data and the smoothing operator used in
the zero-level determination. We therefore assign a residual
uncertainty to this value as described by \citet{CG02_01}. Fourth, the
starting point of the current analysis are the calibrated TOD as
provided by the DIRBE team. This process itself has uncertainties both
in terms of absolute calibration and baseline determination as listed
in Table~1 of \citet{hauser1998}. The baseline is a linear term, and
we therefore propagate this directly as provided. However, the gain
uncertainty is a multiplicative value in units of percent, and we
therefore multiply those uncertainties with our best-fit monopole
values for each channel before adding all terms together in
quadrature. For channels without a positive detection, we define the
upper 95\,\% confidence limit as the sum of the posterior mean value
and two times the total uncertainty.

The results from these calculations are summarized in
Table~\ref{tab:CIB_monopole}, both in terms of individual uncertainty
contributions and measurements and upper limits. The final
\cosmoglobe\ DR2 results are listed in column (j), while the
corresponding constraints from \citet{hauser1998} are reproduced in
column (h). As a simple validation test, column (i) lists constraints
that are derived directly from the monopole Gibbs samples, $m_{\nu}$,
and these are therefore based on a less conservative masking procedure
than the main results. Overall, the results from these two methods are
very similar, and this illustrates that the final results are not
strongly dependent on algorithmic post-processing choices.

Considering the individual contributions to the error budget, we see
that different effects dominate for different channels. For instance,
the posterior uncertainties dominate at 12 and 25$\,\mu\mathrm{m}$,
while at 1.25--3.5$\,\mu\mathrm{m}$ the systematic half-mission
uncertainties dominate. The ultimate goal is that the statistical term
should be the largest factor, and the fact that it not yet is for
several channels indicates that the DIRBE data still have additional
constraining power that can be released through further analysis.

\subsection{Comparison with previous results}

Figure~\ref{fig:CIB_monopoles} compares the final \cosmoglobe\ DR2
constraints with selected previously published results, as well as
with a few representative theoretical models. First, the original
constraints by \citet{hauser1998} are shown as orange markers, eight
of which are upper limits and two are positive detections. In
contrast, our analysis has resulted in several new point estimates as
compared to the original analysis, including between 1.25 and
3.5$\,\mu$m and at 100$\,\mu$m, while for the four channels spanning
4.9 and 60$\,\mu\mathrm{m}$ our limits are generally a factor of two
to eight times stronger than the previous results. We also note that
for the 140 and 240$\,\mu\mathrm{m}$ channels, where
\citet{hauser1998} did report positive detections, our values lower by
64\,\% and 56\,\% than the official DIRBE results, respectively. We
interpret this as being due to better zodiacal light modeling in
\cosmoglobe\ DR2, and conclude that the original estimates were biased
high by 2--$3\,\sigma$.

Shortly after the release of the DIRBE analysis, several authors
reanalyzed the DIRBE ZSMA maps together with complementary external
data sets \citep[e.g.,][]{wright:2000,wright:2001}, conceptually
similar to what is done in the current \cosmoglobe\ DR2 release. As
two concrete examples, the red points in Fig.~\ref{fig:CIB_monopoles}
show the results obtained by \citet{cambresy:2001} when combining
DIRBE ZSMA at 1.25 and 2.2$\,\mu\mathrm{m}$ with 2MASS measurements,
while the blue point shows the result derived by \citet{gorjian:2000}
at 3.5$\,\mu\mathrm{m}$ when combining with dedicated follow-up
observations of a $2^{\circ}\times2^{\circ}$ dark spot near the North
Galactic Pole. While the latter measurement agrees very well with our
measurements, the two former points are higher by 25\,\% and 53\,\%,
respectively, or 1 and $2\,\sigma$. The other colored points in
Fig.~\ref{fig:CIB_monopoles} correspond to a selection of more recent
measurements with other probes. For reference, the continuous line
shows the expected contribution of galaxies from redshifts 0--6 as
estimated by \citet{finke2022}.

\section{Conclusions}
\label{sec:conclusions}

In this work, we have derived improved constraints on the CIB monopole spectrum as observed by DIRBE. This was achieved through global end-to-end Bayesian sampling, in which the monopoles were sampled jointly with both zodiacal and Galactic emission and instrumental parameters. The reprocessed DIRBE maps, as presented by \citet{CG02_01}, have been improved from the legacy processing in large part due to improved zodiacal dust modeling \citep{CG02_02}, deeper stellar modeling \citep{CG02_04}, and a novel thermal dust model \citep{CG02_05}. Notably, the presence of excess radiation that appears static in solar-centric coordinates was identified and removed in the 4.9, 12, 25, and $60\,\mathrm{\mu m}$ bands.

In the current paper we have analyzed a set of data-minus-model residual maps that resulted from this process, which ideally should contain only CIB signal and instrumental noise. Overall, our monopole estimates derived from these maps are significantly lower than those derived from the legacy DIRBE maps. This is true across the entire wavelength band, even at those channels where significant detections were already reported in the literature. Based on the results presented by \citet{CG02_01} and \citet{CG02_02}, we interpret this primarily as improved zodiacal light modeling in the current processing. We anticipate that these improved results and products will have non-trivial implications for many astrophysical and cosmological analyses that were based on the original DIRBE measurements, and stronger limits may now be imposed on a wide range of physical effects.

This progress has been enabled primarily by two defining features of the \cosmoglobe\ framework. The first of these is simply joint analysis of multiple complementary state-of-the-art experiments. In the case of DIRBE, the combination of \Planck\ HFI, WISE and \Gaia\ data has established a new view of emission from the Milky Way in the form of a single sky model that spans $100\,\mathrm{GHz}$ to $1\,\mathrm{\mu m}$, and this has in turn allowed a deeper mapping of the zodiacal light emission than previously possible. The second defining feature of \cosmoglobe\ is the use of modern statistical methods and computing power, which enables global end-to-end modeling of full time-ordered data sets. Intuitively speaking, fitting all parameters at once leads to better estimates of each parameter individually, and such global modeling was simply not computationally feasible when the original DIRBE analysis was performed in the 1990's.

Despite these improvements, there are still outstanding problems and degeneracies that cannot be broken with the data used in this study. The most important of these is the existence of solar-centric excess radiation in the wavelength channels between 4.9, 12, 25, and 60\,$\mathrm{\mu m}$. A natural next step towards understanding this is to establish a detailed straylight model for the DIRBE instrument, for instance using GRASP \citep{grasp}. If a detailed physical optics analysis excludes a straylight-based explanation, more complicated zodiacal light models must be considered. Irrespective of the origin of this effect, the addition of time-ordered data from other experiments at similar frequencies and with complementary scan strategies can be used to better determine the 3D structure and absolute brightness of zodiacal dust. In particular, IRAS \citep{boggess92} created nearly full-sky maps at 12, 25, 60, and 100 $\mathrm{\mu m}$ with resolution between 0.5$\arcm$ and $2\arcm$ FWHM. A full end-to-end joint analysis of IRAS and DIRBE will leverage the unique properties of both datsets, and enable robust characterization of the CIB monopole spanning the entire infrared spectrum. Similarly, other full-sky experiments, including the AKARI \citep{murakami:2007} and the upcoming SPHEREx \citep{dore:2014} satellites, will be essential for determining the three-dimensional structure of the zodiacal dust and determining the spectrum of the CIB, and we argue that all of these should ideally be analyzed jointly within a common end-to-end framework like \Cosmoglobe.

\begin{acknowledgements}
  We thank Tony Banday, Johannes Eskilt, Dale Fixsen, Ken Ganga, Paul
  Goldsmith, Shuji Matsuura, Sven Wedemeyer, and Janet Weiland for useful suggestions
  and guidance.  The current work has received funding from the
  European Union’s Horizon 2020 research and innovation programme
  under grant agreement numbers 819478 (ERC; \textsc{Cosmoglobe}),
  772253 (ERC; \textsc{bits2cosmology}), and 101007633 (MSCA;
  \textsc{CMBInflate}).  Some of the results in this paper have been
  derived using healpy \citep{Zonca2019} and the HEALPix
  \citep{healpix} package.  We acknowledge the use of the Legacy
  Archive for Microwave Background Data Analysis (LAMBDA), part of the
  High Energy Astrophysics Science Archive Center
  (HEASARC). HEASARC/LAMBDA is a service of the Astrophysics Science
  Division at the NASA Goddard Space Flight Center. This publication
  makes use of data products from the Wide-field Infrared Survey
  Explorer, which is a joint project of the University of California,
  Los Angeles, and the Jet Propulsion Laboratory/California Institute
  of Technology, funded by the National Aeronautics and Space
  Administration. This work has made use of data from the European
  Space Agency (ESA) mission {\it Gaia}
  (\url{https://www.cosmos.esa.int/gaia}), processed by the {\it Gaia}
  Data Processing and Analysis Consortium (DPAC,
  \url{https://www.cosmos.esa.int/web/gaia/dpac/consortium}). Funding
  for the DPAC has been provided by national institutions, in
  particular the institutions participating in the {\it Gaia}
  Multilateral Agreement.
\end{acknowledgements}

\bibliographystyle{aa}
\bibliography{Planck_bib,CG_bibliography}

\end{document}

%% file: Planck.tex
\def\setsymbol#1#2{\expandafter\def\csname #1\endcsname{#2}}
\def\getsymbol#1{\csname #1\endcsname}

\def\Planck{\textit{Planck}}

\newbox\tablebox    \newdimen\tablewidth
\def\leaderfil{\leaders\hbox to 5pt{\hss.\hss}\hfil}

\def\endPlancktablewide{\tablewidth=\textwidth 
    $$\hss\copy\tablebox\hss$$
    \vskip-\lastskip\vskip -2pt}
\def\tablenote#1 #2\par{\begingroup \parindent=0.8em
    \abovedisplayshortskip=0pt\belowdisplayshortskip=0pt
    \noindent
    $$\hss\vbox{\hsize\tablewidth \hangindent=\parindent \hangafter=1 \noindent
    \hbox to \parindent{$^#1$\hss}\strut#2\strut\par}\hss$$
    \endgroup}
\def\doubleline{\vskip 3pt\hrule \vskip 1.5pt \hrule \vskip 5pt}

\def\L2{\ifmmode L_2\else $L_2$\fi}

\def\DeltaT{\ifmmode \Delta T\else $\Delta T$\fi}
\def\deltat{\ifmmode \Delta t\else $\Delta t$\fi}
\def\fknee{\ifmmode f_{\rm knee}\else $f_{\rm knee}$\fi}
\def\Fmax{\ifmmode F_{\rm max}\else $F_{\rm max}$\fi}
\def\solar{\ifmmode{\rm M}_{\mathord\odot}\else${\rm M}_{\mathord\odot}$\fi}
\def\Msolar{\ifmmode{\rm M}_{\mathord\odot}\else${\rm M}_{\mathord\odot}$\fi}
\def\Lsolar{\ifmmode{\rm L}_{\mathord\odot}\else${\rm L}_{\mathord\odot}$\fi}
\def\inv{\ifmmode^{-1}\else$^{-1}$\fi}
\def\mo{\ifmmode^{-1}\else$^{-1}$\fi}
\def\sup#1{\ifmmode ^{\rm #1}\else $^{\rm #1}$\fi}
\def\expo#1{\ifmmode \times 10^{#1}\else $\times 10^{#1}$\fi}
\def\,{\thinspace}
\def\lsim{\mathrel{\raise .4ex\hbox{\rlap{$<$}\lower 1.2ex\hbox{$\sim$}}}}
\def\gsim{\mathrel{\raise .4ex\hbox{\rlap{$>$}\lower 1.2ex\hbox{$\sim$}}}}

\def\simprop{\mathrel{\raise .4ex\hbox{\rlap{$\propto$}\lower 1.2ex\hbox{$\sim$}}}}
\def\deg{\ifmmode^\circ\else$^\circ$\fi}
\def\pdeg{\ifmmode $\setbox0=\hbox{$^{\circ}$}\rlap{\hskip.11\wd0 .}$^{\circ}
          \else \setbox0=\hbox{$^{\circ}$}\rlap{\hskip.11\wd0 .}$^{\circ}$\fi}
\def\arcs{\ifmmode {^{\scriptstyle\prime\prime}}
          \else $^{\scriptstyle\prime\prime}$\fi}
\def\arcm{\ifmmode {^{\scriptstyle\prime}}
          \else $^{\scriptstyle\prime}$\fi}
\newdimen\sa  \newdimen\sb
\def\parcs{\sa=.07em \sb=.03em
     \ifmmode \hbox{\rlap{.}}^{\scriptstyle\prime\kern -\sb\prime}\hbox{\kern -\sa}
     \else \rlap{.}$^{\scriptstyle\prime\kern -\sb\prime}$\kern -\sa\fi}
\def\parcm{\sa=.08em \sb=.03em
     \ifmmode \hbox{\rlap{.}\kern\sa}^{\scriptstyle\prime}\hbox{\kern-\sb}
     \else \rlap{.}\kern\sa$^{\scriptstyle\prime}$\kern-\sb\fi}
\def\ra[#1 #2 #3.#4]{#1\sup{h}#2\sup{m}#3\sup{s}\llap.#4}
\def\dec[#1 #2 #3.#4]{#1\deg#2\arcm#3\arcs\llap.#4}
\def\deco[#1 #2 #3]{#1\deg#2\arcm#3\arcs}
\def\rra[#1 #2]{#1\sup{h}#2\sup{m}}

\def\dots{\relax\ifmmode \ldots\else $\ldots$\fi}
\def\WHzsr{\ifmmode $W\,Hz\mo\,sr\mo$\else W\,Hz\mo\,sr\mo\fi}
\def\mHz{\ifmmode $\,mHz$\else \,mHz\fi}
\def\GHz{\ifmmode $\,GHz$\else \,GHz\fi}
\def\mKs{\ifmmode $\,mK\,s$^{1/2}\else \,mK\,s$^{1/2}$\fi}
\def\muKs{\ifmmode \,\mu$K\,s$^{1/2}\else \,$\mu$K\,s$^{1/2}$\fi}
\def\muKRJs{\ifmmode \,\mu$K$_{\rm RJ}$\,s$^{1/2}\else \,$\mu$K$_{\rm RJ}$\,s$^{1/2}$\fi}
\def\muKHz{\ifmmode \,\mu$K\,Hz$^{-1/2}\else \,$\mu$K\,Hz$^{-1/2}$\fi}
\def\MJysr{\ifmmode \,$MJy\,sr\mo$\else \,MJy\,sr\mo\fi}
\def\MJysrmK{\ifmmode \,$MJy\,sr\mo$\,mK$_{\rm CMB}\mo\else \,MJy\,sr\mo\,mK$_{\rm CMB}\mo$\fi}
\def\microns{\ifmmode \,\mu$m$\else \,$\mu$m\fi}

\def\muK{\ifmmode \,\mu$K$\else \,$\mu$\hbox{K}\fi}
\def\microK{\ifmmode \,\mu$K$\else \,$\mu$\hbox{K}\fi}
\def\muW{\ifmmode \,\mu$W$\else \,$\mu$\hbox{W}\fi}
\def\kms{\ifmmode $\,km\,s$^{-1}\else \,km\,s$^{-1}$\fi}
\def\kmsMpc{\ifmmode $\,\kms\,Mpc\mo$\else \,\kms\,Mpc\mo\fi}

\providecommand{\sorthelp}[1]{}

%% file: authors.tex
\author{\small
D.~J.~Watts\inst{\ref{uio}}\thanks{Corresponding author: D.~Watts; \url{duncan.watts@astro.uio.no}}
\and
M.~Galloway\inst{\ref{uio}}
\and
E.~Gjerl\o w\inst{\ref{uio}}
\and
M.~San\inst{\ref{uio}}
\and
R.~Aurlien\inst{\ref{uio}}
\and
A.~Basyrov\inst{\ref{uio}}
\and
M.~Brilenkov\inst{\ref{uio}}
\and
H.~K.~Eriksen\inst{\ref{uio}}
\and
U.~Fuskeland\inst{\ref{uio}}
\and
L.~T.~Hergt\inst{\ref{ubc}}
\and
D.~Herman\inst{\ref{uio}}
\and
H.~T.~Ihle\inst{\ref{uio}}
\and
J.~G.~S.~Lunde\inst{\ref{uio}}
\and
S.~K.~Næss\inst{\ref{uio}}
\and
N.-O.~Stutzer\inst{\ref{uio}}
\and
H.~Thommesen\inst{\ref{uio}}
\and
I.~K.~Wehus\inst{\ref{uio}}
}
\institute{\small
Institute of Theoretical Astrophysics, University of Oslo, Blindern, Oslo, Norway\label{uio}
\and
Department of Physics and Astronomy, University of British Columbia, 6224 Agricultural Road, Vancouver BC, V6T1Z1, Canada\label{ubc}
}